\documentclass{JHEP}

\usepackage{amssymb}

\usepackage{amsfonts}

\usepackage{amsbsy}

\usepackage{epsfig}

\newcommand{\im}{\mathop{\rm Im}\nolimits}
\newcommand{\re}{\mathop{\rm Re}\nolimits}
\newcommand{\NP}[1]{Nucl.\ Phys.\ {\bf #1}}

\renewcommand{\cal}{\mathcal}
\newcommand{\sfrac}[2]{{\textstyle \frac{#1}{#2}}}

\def\IZ{\mathbb{Z}}
\def\IR{\mathbb{R}}

\def\sx{{\bf x}}

\def\CK{{\cal K}}

\def\N{{\cal N}}
\def\bb{\bar{b}}
\def\bz{\bar{z}}

\def\bpartial{\bar{\partial}}

\def\s*{\boldsymbol{*}}

\def\MG{\mathrm{MeijerG}}

\newcommand{\be}{\begin{equation}}
\newcommand{\ee}{\end{equation}}
\newcommand{\bea}{\begin{eqnarray}}
\newcommand{\eea}{\end{eqnarray}}
\newcommand{\nn}{\nonumber}
\def\C{{\mathbb C}}
\def\Om{\Omega}
\def\Omn{\tilde{\Omega}}

\title{Split attractor flows and the spectrum of BPS D-branes on the Quintic}

\author{Frederik Denef, Brian Greene and Mark Raugas \\ Department of Mathematics and Physics,
Columbia University \\ New York, NY 10027, USA\\ \email{denef,
greene, raugasm@math.columbia.edu}}

\preprint{ CU-TP-1001\\{\tt hep-th/0101135}}

\abstract{We investigate the spectrum of type IIA BPS D-branes on
the quintic from a four dimensional supergravity perspective and
the associated split attractor flow picture. We obtain some very
concrete properties of the (quantum corrected) spectrum, mainly
based on an extensive numerical analysis, and to a lesser extent
on exact results in the large radius approximation. We predict the
presence and absence of some charges in the BPS spectrum in
various regions of moduli space, including the precise location of
the lines of marginal stability and the corresponding decay
products. We explain how the generic appearance of multiple basins
of attraction is due to the presence of conifold singularities and
give some specific examples of this phenomenon. Some interesting
space-time features of these states are also uncovered, such as a
nontrivial, moduli independent lower bound on the area of the core
of arbitrary BPS solutions, whether they are black holes, empty
holes, or more complicated composites.}

\begin{document}

\section{Introduction}\label{sec:intro}
\setcounter{equation}{0}

Type II string theory compactified on a Calabi-Yau manifold has a
rich and very nontrivial spectrum of BPS states, obtained by
wrapping D-branes around various supersymmetric cycles in the
compactification manifold. Over the past years, several approaches
have been developed to analyze the notorious problem of existence
and decay of these objects. Essentially there are two
complementary viewpoints --- a common theme in D-brane physics:
the microscopic D-brane picture and the macroscopic supergravity
description. The former has been studied by far the most, as it
provides a direct, powerful and extensive framework, firmly rooted
in geometry and string theory. A partial list of references is
\cite{RS,BDLR,D,DR,Sche,BruScho,DFR,DFR2,DD,FM,Lerche,KachruLectures,OPW,Dcat,T,GJ2,
mayr,tomeu,joyce,KM,GJ,AV}.
The alternative approach, based on the four dimensional low energy
effective $\N=2$ supergravity theory, has received considerably
less attention, in part because at first sight this effective
theory seemed much too simple to be able to capture the
intricacies of the D-brane spectrum. To a large extent, this is
true, though a number of intriguing surprises were uncovered,
starting with the work of \cite{M}, where BPS black hole solutions
and the associated attractor mechanism of \cite{FKS} were studied,
and it was found that existence of these solutions for given
charge and moduli is quite nontrivial and linked to some deep
results in microscopic D-brane physics and arithmetic, suggesting
a correspondence between existence of BPS black hole solutions and
BPS states in the full string theory. This conjecture turned out
to fail in a rather mysterious way in a number of cases
\cite{D,branessugra}, but it was soon realized how to fix this:
more general, non-static (but stationary) multicenter solutions
have to be considered, and corresponding to those, more general
attractor flows, the so-called \emph{split} attractor flows
\cite{branessugra,montreal}. Thus, an unexpectedly rich structure
of solutions emerged, providing a low energy picture of many
features of the BPS spectrum, including bound states, intrinsic
angular momentum, decay at marginal stability, and a number of
stability criteria tantalizingly similar to those obtained from
microscopic brane physics and pure mathematics. Moreover, further
evidence was found for the correspondence between (split)
attractor flows and stringy BPS states, which might even extend
beyond the supergravity regime.

In this paper, we further explore this approach, focusing on the
well-known and widely studied example of type IIA theory
compactified on the quintic Calabi-Yau (or IIB on its mirror)
\cite{cand,BDLR}, mainly on the basis of an extensive numerical
analysis. We give various examples of the sort of results that can
be obtained using the split flow picture, predict presence and
absence of some charges in various regions of moduli space
(including precise lines of marginal stability and corresponding
decay products), elucidate the appearance of multiple basins of
attraction and give some examples of this phenomenon, compare our
numerical results for the fully instanton-corrected theory with
analytical results in the large radius approximation, obtain an
interesting lower bound on the area of all BPS objects in the
supergravity theory, and briefly discuss some extentions to
non-BPS states and the zoo of near horizon split flows. Our main
conclusion is that surprisingly much can be learned from the
supergravity picture, in a very concrete way, though fundamental
insight in the spectrum is still more likely to come from the
microscopic picture.

The outline of the paper is as follows. In section 2, we review
the construction of stationary BPS solutions of four dimensional
$\N=2$ supergravity (including the enhan\c{c}on \cite{enhancon}
related empty hole) and their relation to split flows, and give
some comments on the validity of the four dimensional supergravity
approximation. In section 3, we explain how one can obtain (split)
attractor flow spectra, based on a number of existence criteria,
and we show how the presence of singularities generically induces
multiple basins of attraction, and how the split flow picture
avoids a clash with microscopic entropy considerations. Section 4
reviews the essentials of compactification on the quintic. In
section 5, we outline the practical strategies we followed for
computing attractor flows, which we apply in section 6 to a broad
analysis of the quintic. More precisely, in section 6.2, some
features of the single flow spectrum around the Gepner point are
analyzed, including a screening of a large number of candidate BPS
states, supporting physical expectations of discreteness of the
BPS spectrum. In 6.3, we give an example of a charge that exists
as a BPS black hole at large radius, but decays when the moduli
are varied towards the Gepner point, where it is absent from the
spectrum, providing a nice qualitative distinction between large and small radius physics.  Section 6.4 gives an example of an interesting bound
state of two black holes at large radius that does not exist as a
single black hole [but does have yet another composite
realization, without (regular, four dimensional) black hole
constituents] and decays on its way to the Gepner point. Section
6.5 gives an example of the mulitple basin phenomenon. In 6.6 we
have a look at exact results in the large radius approximation,
and compare this with numerical results for the interesting
example of D6-D2 states. A puzzle related to the stability of
solutions in the presence of conifold singularities is raised in
section 6.7, but not conclusively resolved, though we suggest some
possible ways out. In 6.8 we go back to spacetime properties of
the solutions, and find that they all satisfy a certain area
bound, which we explicitly compute. Finally, in 6.9, we briefly
comment on multicenter configurations in the near horizon region
of a black hole, and find that there are many more possibilities
here than in asymptotically flat space. We end with our
conclusions and some discussion in section 7.

\section{BPS solutions of 4d N=2 supergravity and split
attractor flows}\label{sec:BPS} \setcounter{equation}{0}

\subsection{Special geometry of type IIB Calabi-Yau compactifications} \label{sec2}

For concreteness, we will assume that the four dimensional ${\cal
N}=2$ supergravity theory is obtained from a compactification of
type IIB string theory on a Calabi-Yau 3-fold $X$. This theory has
$n_v = h^{1,2}(X)$ massless abelian vector multiplets and $n_h =
h^{1,1}(X)+ 1$ massless hypermultiplets. The hypermultiplet fields
will play no role here and are set to constant values.

The vector multiplet scalars are given by the complex structure
moduli of $X$, and the lattice of electric and magnetic charges is
identified with $H^3(X,\IZ)$, the lattice of integral harmonic
$3$-forms on $X$: after a choice of symplectic basis
${\alpha^I,\beta_I}$ of $H^3(X,\IZ)$, a D3-brane wrapped around a
cycle Poincar\'e dual to $\Gamma \in H^3(X,\IZ)$ has electric and
magnetic charges equal to its components with respect to this
basis.

The geometry of the vector multiplet moduli space, parametrized by
$n_v$ coordinates $z^a$, is special K\"ahler~\cite{SG}. The
(positive definite) metric
\begin{equation} \label{SKmetric}
 g_{a\bb} = \partial_a \bpartial_{\bb} \CK
\end{equation}
is derived from the K\"ahler potential
\begin{equation} \label{kahlerpotential}
\CK = - \ln \left( i \int_X \Om \wedge \bar{\Om} \right),
\end{equation}
where $\Om$ is the holomorphic $3$-form on $X$, depending
holomorphically on the complex structure moduli. It is convenient
to introduce also the {\em normalized} 3-form\footnote{In
\cite{branessugra,montreal}, the holomorphic 3-form was denoted as
$\Omega_0$, and the normalized one as $\Omega$.}
\begin{equation} \label{Omdef}
\Omn  \equiv e^{\CK/2} \, \Om\,.
\end{equation}
The ``central charge'' of $\Gamma \in H^3(X,\IZ)$ is given by
\begin{equation} \label{Zdef}
Z(\Gamma) \equiv \int_X \Gamma \wedge \Omn \equiv \int_\Gamma
\Omn\,,
\end{equation}
where we denoted, by slight abuse of notation, the cycle
Poincar\'e dual to $\Gamma$ by the same symbol $\Gamma$. Note that
$Z(\Gamma)$ has a nonholomorphic dependence on the moduli through
the K\"{a}hler potential.

The (antisymmetric, topological, moduli independent)
\emph{intersection product} is defined as:
\begin{equation} \label{intproddef}
\langle \Gamma_1,\Gamma_2 \rangle = \int_X \Gamma_1 \wedge
\Gamma_2 = \int_{\Gamma_1} \Gamma_2 = \#(\Gamma_1 \cap
\Gamma_2)\,.
\end{equation}
With this notation, we have for a symplectic basis $\{
\alpha^I,\beta_I \}$ by definition $\langle \alpha^I,\beta_J
\rangle = \delta^I_J$, so for $\Gamma_i = q_i^I \beta_I - p_{i,I}
\alpha^I$, we have $\langle \Gamma_1,\Gamma_2 \rangle = q_1^I
p_{2,I} - p_{1,I} q_2^I$. This is nothing but the
Dirac-Schwinger-Zwanziger symplectic inner product on the
electric/magnetic charges. Integrality of this product is
equivalent to Dirac charge quantization.

\subsection{BPS configurations} \label{BPSeom}

\subsubsection{Single charge type: single flows} \label{sec:single}

Static, spherically symmetric BPS configurations \cite{FKS,FGK,attrsusy,attrsols}
with charge $\Gamma \in H^3(X,\IZ)$ at the origin of space have a
spacetime metric of the form
\begin{equation} \label{staticmetric}
 ds^2 = - e^{2U} dt^2 + e^{-2 U} dx^i
dx^i \,,
\end{equation}
with $U$ a function of the radial coordinate distance
$r=|\bf{x}|$, or equivalently of the inverse radial coordinate
$\tau=1/r$. We will take space to be asymptotically flat, with
$U_{\tau=0}=0$. The BPS equations of motion for $U(\tau)$ and the
moduli $z^a(\tau)$ are:
\begin{eqnarray}
 \partial_\tau U &=& - e^U |Z| \,,\label{at1} \\
 \partial_\tau z^a &=& -2 e^U g^{a\bb} \, \bpartial_{\bb} |Z|\,, \label{at2}
\end{eqnarray}
where $Z=Z(\Gamma)$ is as in (\ref{Zdef}) and $g_{a\bb}$ as in
(\ref{SKmetric}). The electromagnetic field is given algebraically
and in closed form in terms of the solutions of these flow
equations, but we will not need the explicit expression here.

\FIGURE[t]{\centerline{\epsfig{file=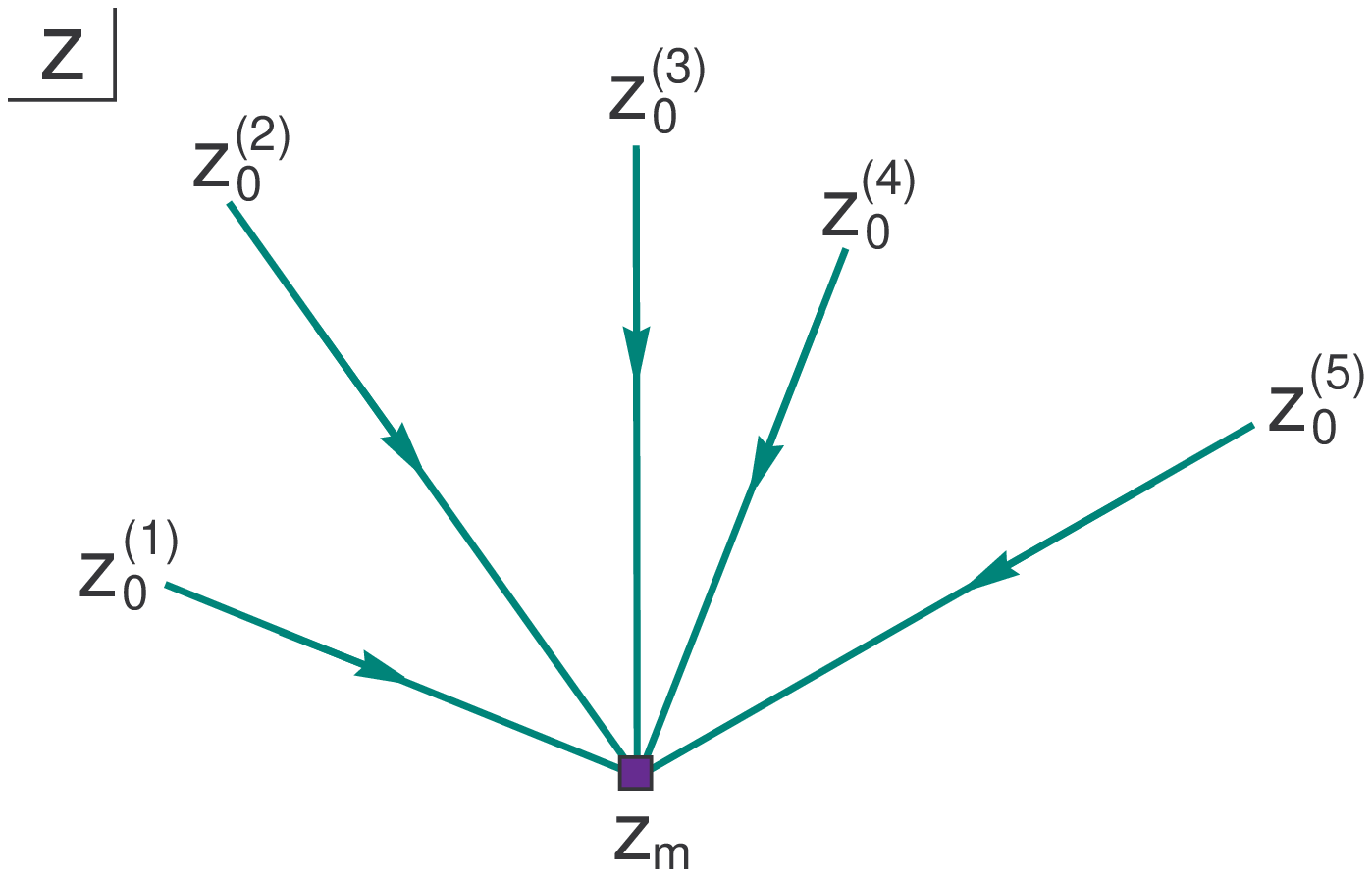,height=5.5cm}}
\caption{Sketch of some single charge attractor flows in moduli
space, for different values $z_0^{(i)}$ of the moduli at spatial
infinity. The attractor point is indicated by
$z_m$.}\label{flowsketch}}

An alternative form of the equations is:
\begin{equation} \label{bps3}
2 \, \partial_\tau [ e^{-U} \im (e^{-i \alpha} \Omn )] = -
\Gamma\,,
\end{equation}
where $\alpha = \arg Z$, which can be shown to be the phase of the
conserved supersymmetry \cite{M}. Note that this nice compact
equation actually has $2 n_v + 2$ real components, corresponding
to taking intersection products with the $2 n_v + 2$ elements of a
basis $\{C_L\}_L$ of $H^3(X,\IZ)$:
\begin{equation} \label{bpscomp}
2 \, \partial_\tau [ e^{-U} \im ( e^{-i \alpha} Z(C_L) )] = -
\langle C_L , \Gamma \rangle \,,
\end{equation}
One component is redundant, since taking the intersection product
of (\ref{bps3}) with $\Gamma$ itself produces trivially $0=0$.
This leaves $2 n_v + 1$ independent equations, matching the number
of real variables $\{U,\re z^a,\im z^a\}$.

Since the right hand side of (\ref{bpscomp}) consists of
$\tau$-independent integer charges, (\ref{bps3}) integrates to
\begin{equation} \label{integrated}
 2\, e^{-U} \im [e^{-i \alpha} \Omn]= -\Gamma \, \tau +
 2  \im [e^{-i \alpha} \Omn]_{\tau=0}.
\end{equation}
This solves in principle the equations of motion. Of course,
finding the explicit flows in moduli space from~(\ref{integrated})
requires inversion of the periods to the moduli, which in general
is not feasible analytically. For this paper, which studies the
case of the quintic Calabi-Yau for arbitrary values of the moduli,
we developed some numerical approaches to tackle this problem.

Generalization to the multicenter BPS configurations with {\em
identical} charges $\Gamma$ at locations $\sx_p$ (arbitrary and
possibly coinciding) is straightforward: one just replaces
$\tau=1/|\sx|$ by $\sum_p 1/|\sx-\sx_p|$. Thus, the flow in moduli
space will remain the same, only its spacetime parametrization
changes.

It was observed in \cite{M} that the attractor flows in moduli
space given by the BPS equations do not always exist. While
solutions to (\ref{at1})-(\ref{at2}) generically do exist for a
finite range of $\tau$ (starting from spatial infinity $\tau=0$),
they can break down before $\tau=\infty$ is reached. This can be
seen as follows (see e.g.\ \cite{M,branessugra,montreal} for more
details). The BPS equations imply that, away from a singular point
or a critical point of $|Z|$,
\begin{equation} \label{Mdecr}
\partial_\tau
|Z| = -4 \, e^{U} g^{a\bb} \,
\partial_a |Z| \, \bpartial_{\bb} |Z| < 0 \, ,
\end{equation}
so along a flow, $|Z|$ is a decreasing function, converging to a
local minimum, the so-called attractor point (see fig.\
\ref{flowsketch}). Three cases are distinguished \cite{M},
depending on the value $|Z_m|$ and the position $z_m$ of this
minimum in moduli space:
\begin{enumerate}
\item {\em $|Z_m| \neq 0$}: the flow exists all the
way up to $\tau=\infty$ and the solution exists as a regular BPS
black hole, with $\mathrm{AdS}_2 \times S^2$ near horizon geometry
and horizon area $A=4 \pi |Z_m|^2$. Note that the horizon moduli
$z_m$ are generically\footnote{see however section
\ref{sec:multibasin}} invariant under continuous variations of the
moduli at spatial infinity. The moduli at the horizon satisfy the
so-called attractor equation:
\begin{equation} \label{attreq}
 2 \im(\bar{Z} \Omn) = - \Gamma
\end{equation}

\item {\em $|Z_m|=0$ and $z_m$ is a regular point of moduli space}: the
flow breaks down at finite $\tau$, where the zero of $Z$ is
reached, since at this point, the inequality in (\ref{Mdecr}) does
not make sense. So no BPS solutions exists in this case. This is
compatible with physical expectations, since the existence of a
BPS state with the given charge in a vacuum where $Z$ vanishes
would imply the existence of a massless particle there, which in
turn is expected to create a singularity in moduli space at the
zero, contradicting the assumption of regularity of that point.
Or, from a geometric point of view: if a supersymmetric wrapped
brane exists at the zero, its volume is zero, so we must have a
vanishing cycle in the Calabi-Yau, leading to a singularity in
moduli space.

\item {\em $|Z_m|=0$ and $z_m$ is a singular (or boundary) point in moduli
space}: in this case the arguments of (2) for nonexistence fail,
and indeed well-behaved solutions may exist, like for example the
repulson-resolving \emph{empty hole} solutions of
\cite{branessugra}, arising from flows attracted to a conifold
locus. These correspond to the famous states of \cite{S},
resolving the conifold singularity in string theory.
\end{enumerate}

\subsubsection{Mutually nonlocal charges: split flows}
\label{sec:spliflo}

\FIGURE[t]{\centerline{\epsfig{file=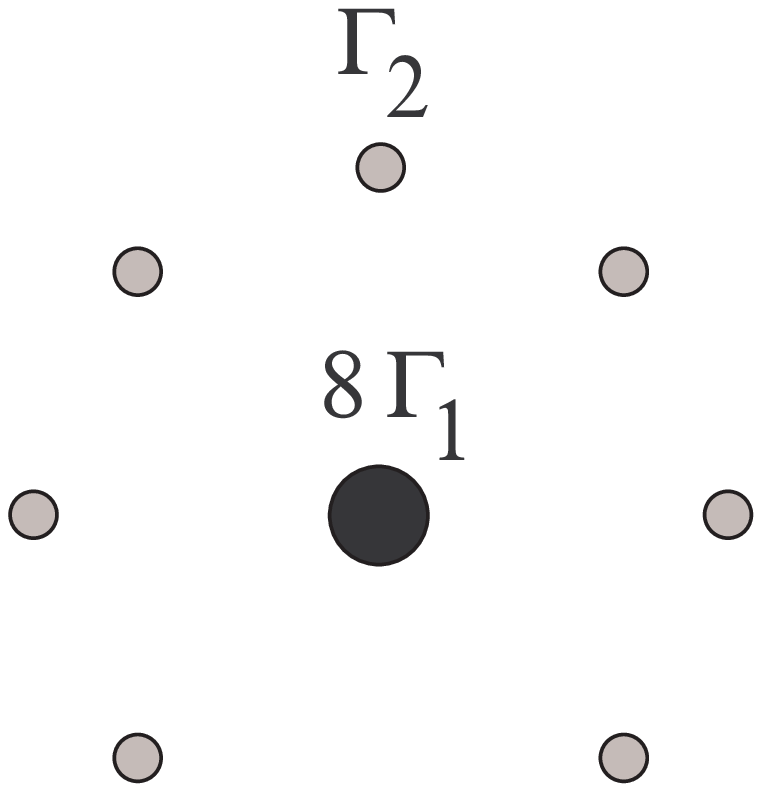,height=5cm}}
\caption{Example of a $(\Gamma_1,\Gamma_2)$-multicenter
configuration with total charge $8 \, \Gamma_1 + 7 \,
\Gamma_2$.}\label{mcconfig}}

It was shown in \cite{branessugra} that for some examples of BPS
states, established by CFT methods to exist in the full string
theory, the corresponding flows in moduli space break down at a
regular zero, making it is necessary to consider more general BPS
solutions, in particular multicenter solutions with mutually
nonlocal charges (fig.\ \ref{mcconfig}). Unfortunately, the BPS
equations \cite{statsugra1,branessugra,CWKM}
--- though formally quite similar to the spherically
symmetric equations --- become substantially more complicated to
solve in this case. This is partly due to the fact that with
mutually nonlocal charges, solutions are in general no longer
static, as they acquire an intrinsic angular momentum (even though
the charge positions are time independent), a fact that is well
known from ordinary Maxwell electrodynamics with magnetically and
electrically charged particles.

\FIGURE[t]{\centerline{\epsfig{file=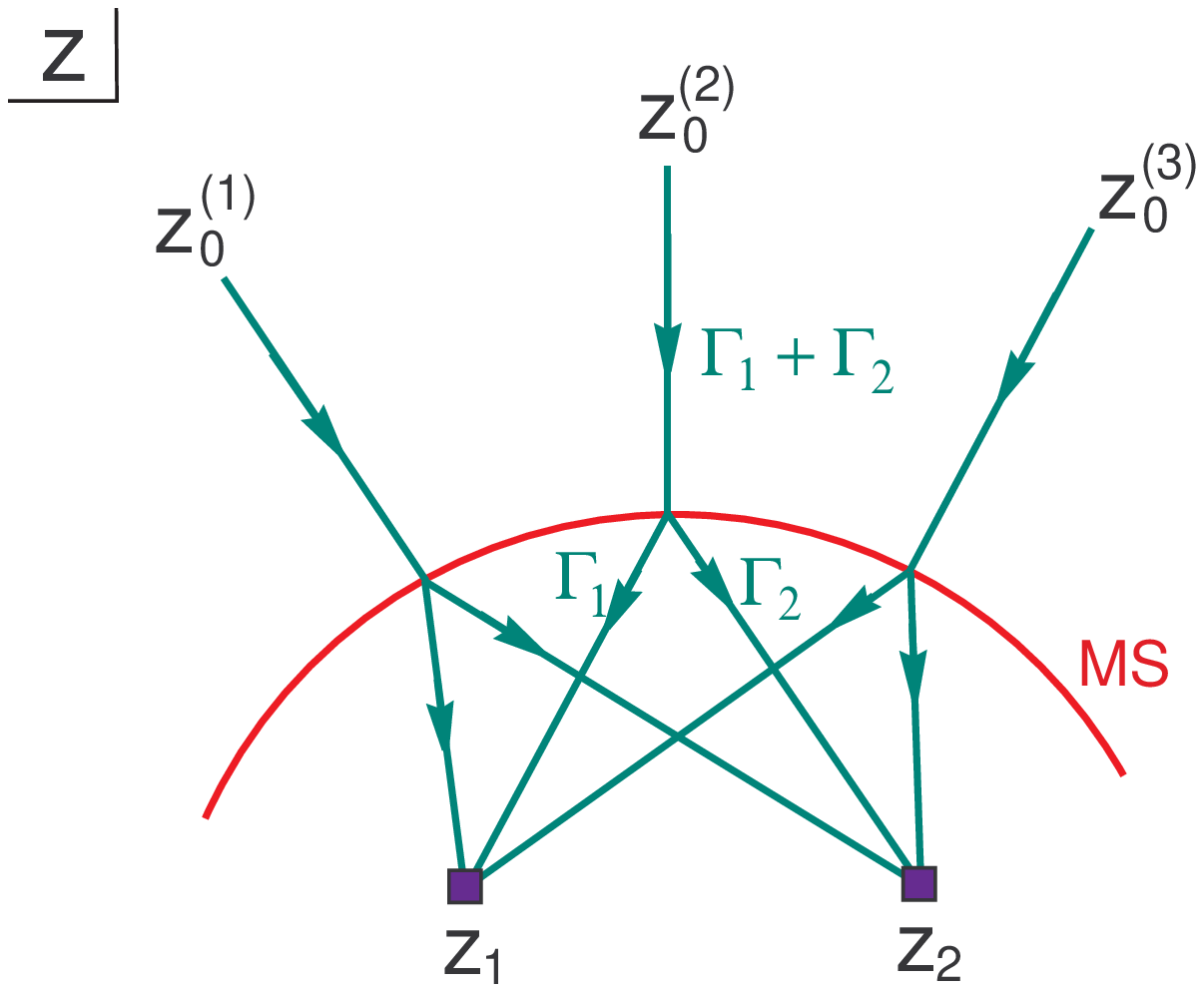,height=5cm}}
\caption{Sketch of some split flows for different moduli values
$z_0^{(i)}$ at spatial infinity, in the case of two different
constituent charges $\Gamma_1$ and $\Gamma_2$, with attractor
points $z_1$ resp.\ $z_2$. The line labeled ``MS'' is (part of)
the $(\Gamma_1,\Gamma_2)$-marginal stability
line.}\label{splitsketch}}

The metric in this case is given by an expression of the form
\begin{equation} \label{stationarymetric}
 ds^2 = - e^{2U} \left(dt + \omega_i \, dx^i\right)^2 + e^{-2 U} dx^i
 dx^i\,,
\end{equation}
and (\ref{integrated}) elegantly generalizes to
\begin{eqnarray}
 2 \, e^{-U} \im [ e^{-i\alpha} \Omn ] &=& H \,, \label{mc1}\\
 \s* d \omega &=& \langle d H,H \rangle \,, \label{mc2}
\end{eqnarray}
with $H(\sx)$ an $H^3(X)$-valued harmonic function (on flat
coordinate space $\IR^3$), and $\s*$ the flat Hodge star operator
on $\IR^3$. For $N$ charges $\Gamma_p$ located at coordinates
$\sx_p$, $p=1,\ldots,N$, in asymptotically flat space, one has:
\begin{equation}
  H = -\sum_{p=1}^N \Gamma_p \, \tau_p  \, \, + \,  2 \im [ e^{-i
  \alpha} \Omn ]_{r=\infty} \,, \label{Haha}
\end{equation}
with $\tau_p=1/|\sx-\sx_p|$.

It was shown in \cite{branessugra} and in more detail in
\cite{montreal}, that such multicenter BPS configurations do
indeed exist, and the existence question in a particular situation
essentially boils down to existence of a corresponding {\em split}
attractor flow, instead of the single flow associated to the
single charge case (see fig.\ \ref{splitsketch}). The endpoints of
the attractor flow branches are the attractor points of the
different charges $\Gamma_p$ ($p=1,\ldots,N$) involved, which are
located at equilibrium positions $\sx_p$ subject to the constraint
\begin{equation} \label{distconstr}
\sum_{q=1}^N \frac{\langle \Gamma_p,\Gamma_q
\rangle}{|\sx_p-\sx_q|} = 2 \, \im [ e^{-i \alpha} Z(\Gamma_p)
]_{r=\infty},
\end{equation}
with $\alpha = \arg Z(\sum_p \Gamma_p)$. For such a multicenter
solution, the image of the moduli fields in moduli space will look
like a ``fattened'' version of the split flow (fig.\
\ref{fattened}). In analogy with the picture arising in the brane
worldvolume description of low energy quantum field theory, one
could imagine space as a ``3-brane'' embedded in moduli space
through the moduli fields $z^a(\sx)$, with the positions of the
charges mapped to their respective attractor points in moduli
space.

\FIGURE[t]{\centerline{\epsfig{file=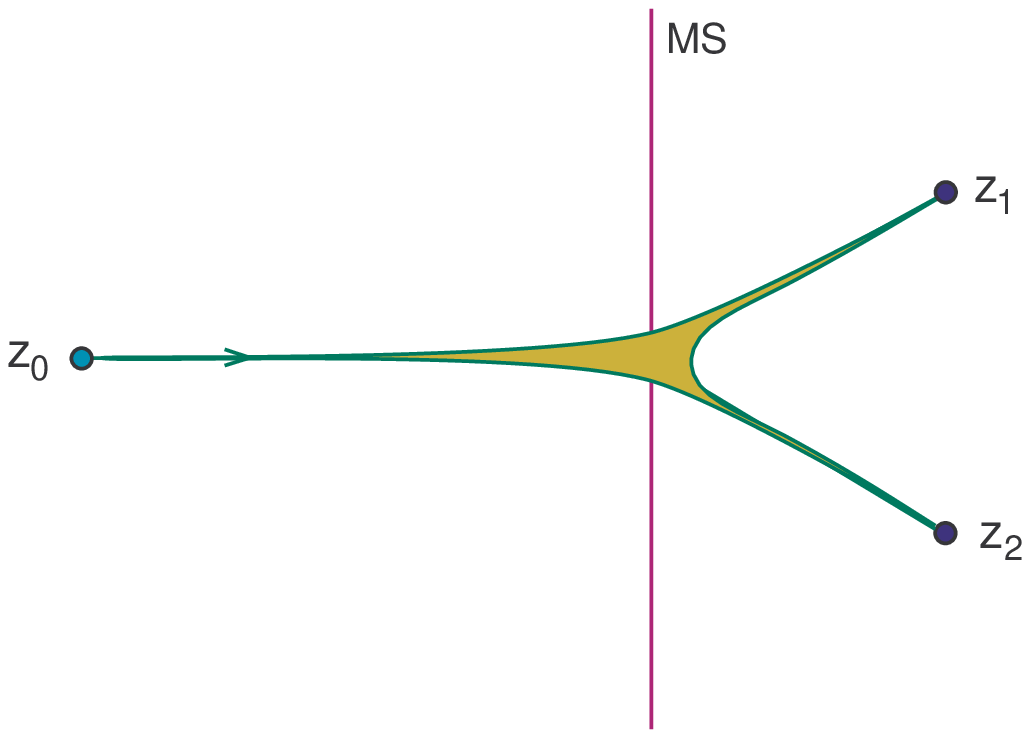,height=5cm}}
\caption{Sketch of the image of $z(\sx)$ in moduli space for a
$(\Gamma_1,\Gamma_2)$-multicenter solution, with attractor points
$z_1$ resp.\ $z_2$, and modulus at spatial infinity $z_0$. The
line labeled ``MS'' is a $(\Gamma_1,\Gamma_2)$-marginal stability
line.}\label{fattened}}

It turns out that the splitting points of the flows have to lie on
a surface of marginal stability in moduli space
\cite{branessugra}; more precisely, a $\Gamma$-flow can only split
in a $\Gamma_1$- and a $\Gamma_2$-flow at a surface of
$(\Gamma_1,\Gamma_2)$-marginal stability, that is, where $\arg
Z(\Gamma_1) = \arg Z(\Gamma_2)$.

We will primarily consider situations with only two different charges
$\Gamma_1$ and $\Gamma_2$ (located in an arbitrary number of
centers), for example a core at the origin of space with $N$
charges $\Gamma_1$ surrounded by a homogeneous ``cloud'' of $N$
charges $\Gamma_2$, constrained by (\ref{distconstr}) to lie on a
sphere of coordinate radius
\begin{equation} \label{rmsformula}
r_{\mathrm{ms}} = N \, \frac{1}{2} \langle \Gamma_1,\Gamma_2
\rangle \left. \frac{|Z_1+Z_2|}{\im(\bar{Z_2} Z_1)}
\right|_{r=\infty} \, .
\end{equation}
Note that when the moduli at infinity approach the surface of
marginal stability, the right hand side of (\ref{rmsformula})
diverges, yielding a smooth decay of the BPS bound state into its
constituents, as could be physically expected. More complicated
split flows can also be considered, with branches splitting
several times. However, many of the features of those more complex
configurations can be understood by iteration of what is known
about split flows with just one split point.

Thus, the main point of this section is that the supergravity BPS
spectrum is essentially given by the spectrum of attractor flows
on moduli space, {\em including} split flows. This will be the
basic starting point for our exploration of the BPS spectrum of
type IIA string theory compactified on the quintic (or its IIB
mirror).

There is one comment to be made here though: composite
configurations involving charges that give rise to empty hole
solutions (when those charges are isolated), such as particles
obtained by wrapping a D3-brane around a vanishing conifold cycle,
seem to be somewhat more subtle than their siblings consisting
exclusively of regular black hole components. In particular, no
explicit truly solitonic\footnote{By ``truly solitonic'' in the
context of effective abelian supergravity theories, we mean a
solution like a BPS black hole or an empty hole, where all mass
can be considered to be located in the four dimensional low energy
fields (so no ``bare'' mass), and the sources' only role is to
generate the required charge.} construction of the former was
given in \cite{montreal}, only an idealized spherical shell
configuration, which requires the addition of a shell of smeared
out wrapped branes with nonvanishing bare mass (whose existence,
strictly speaking, cannot directly be predicted by the
supergravity theory alone). Probably such a construction can be
given with appropriately delocalized charge located on a
``superconducting'' surface, but this might put additional
constraints on the existence of the solution, beyond those implied
by the mere existence of a split flow. We hope to address this
issue elsewhere.

\subsection{Validity of the four dimensional supergravity approximation}

Denote the four dimensional Newton constant by $G_N$. In the IIB
theory, $G_N$ is related to the string coupling constant $g$, the
string scale $l_s$ and the volume $V_X$ of the Calabi-Yau manifold
by $G_N \sim g^2 l_s^8 / V_X$. The four dimensional effective
supergravity description of the IIB string theory can only be
trusted if the characteristic distance scale\footnote{The
curvature scale of the solution will indeed be of this order $L$
if the solution is sufficiently regular \cite{M}. For more
singular solutions, e.g.\ for pure D0-charge, the curvature can
diverge near the singularity, leading to an unavoidable breakdown
of the four dimensional supergravity approximation there.} $L \sim
N \sqrt{G_N}$ of the charge $N$ solution under consideration is
much larger than the string scale $l_s$, the inverse mass
$m^{-1}=|Z|^{-1} \sqrt{G_N}$ of the lightest BPS particle obtained
by wrapping a D3-brane, and the ``size'' $R_X$ of the internal
Calabi-Yau manifold $X$. With ``size'' we mean any relevant linear
dimension of $X$; hence $R_X$ is in part dependent on both the complex
and K\"ahler structure moduli. These dimensions have to be
sufficiently small to justify the four dimensional approximation.

Note that the second and third conditions are dependent on the
complex structure moduli, so for some solutions, it might be
impossible to satisfy this condition everywhere in space, since
the moduli could be driven to values where $|Z| \to 0$ or $R_X \to
\infty$. The former is the case for example for the empty hole
solutions discussed above. So in principle, we should include an
additional light field in the Lagrangian near the core of such
solutions. At large $N$, this presumably would only have the
effect of somewhat smoothing out the solution. It would be
interesting to study this in more detail.

The complication $R_X \to \infty$ occurs typically for charges of
$D0-D2$ type in the IIA picture: the moduli are driven to large
complex structure, where the dimensions of the IIB Calabi-Yau
transverse to the corresponding IIB D3-branes become infinitely
large.\footnote{Note however that the total volume $V_X$ remains
constant, since it does not depend on the complex structure
moduli.}

The physical 4d low energy arguments based on supergravity
considerations we present in this paper are only valid if the
above conditions are met. However, some arguments rely only on
energy conservation considerations starting from the BPS formula,
and since this formula is protected by supersymmetry, those
arguments should also hold outside the supergravity regime. Also,
the conjectured correspondence between BPS states and (split)
attractor flows itself might extend beyond the supergravity
regime. We will refer to this as the strong version of the
conjecture.

\section{Attractor flow spectra}

\subsection{Existence criteria for split flows} \label{sec:existencecrit}

As illustrated in fig.~\ref{splitsketch}, in order for a split
flow $\Gamma \to (\Gamma_1,\Gamma_2)$ to exist, the following two
conditions have to be satisfied:
\begin{enumerate}
 \item The single flow corresponding to the total charge
 $\Gamma = \Gamma_1 + \Gamma_2$,
 starting at the value of the moduli at spatial infinity, has to
 cross a surface of $(\Gamma_1,\Gamma_2)$-marginal stability.
 \item Starting from this crossing point, both the $\Gamma_1$-flow and the
 $\Gamma_2$-flow should exist.
\end{enumerate}
For more complicated split flows (with more split points), these
condition have to be iterated.

A simple necessary condition for condition (1) can be derived from
the integrated BPS equation (\ref{integrated}). Taking the
intersection product of this equation with $\Gamma_1$ gives
\begin{equation}
 \label{imZ1}
2 e^{-U} \im (e^{-i \alpha} Z_1) = - \langle \Gamma_1,\Gamma
\rangle \, \tau  +  2 \, \im (e^{-i \alpha} Z_1)_{\tau=0} \,.
\end{equation}
When $Z_1$ and $Z_2$ are parallel (or anti-parallel), the left
hand side vanishes. On the other hand, since the right hand side
is linear in $\tau$, and $\tau$ has to be positive, this can at
most happen once along the flow, namely iff
\begin{equation} \label{rmscond}
 r_{\mathrm{ms}} \equiv 1/\tau_{\mathrm{ms}} \equiv \frac{1}{2} \langle \Gamma_1,\Gamma_2
 \rangle \left. \frac{|Z_1+Z_2|}{\im(Z_1 \bar{Z_2})}
 \right|_{r=\infty} > 0 \, ,
\end{equation}
\emph{and} the flow does \emph{not} hit a zero before $Z_1$ and
$Z_2$ become (anti-)parallel (when the full flow has a regular
attractor point, the latter is of course automatically satisfied).
Furthermore, only the case where $Z_1$ and $Z_2$ become parallel
(so $|Z|>|Z_1|$ and $|Z|>|Z_2|$) rather than anti-parallel (so
$|Z_1|>|Z|$ or $|Z_2|>|Z|$) gives rise to a split flow. So
(\ref{rmscond}) is a necessary but not sufficient condition for
(1) to hold.

A few simple observations can be made at this point:
\begin{itemize}

\item From the discussion of equation (\ref{rmsformula}) in the
previous section, it follows that this existence condition for a
split flow is just the statement that the radius of separation
between two differently charged source centers is positive. When
the moduli at infinity approach the surface of marginal stability,
this radius diverges, and the configuration decays smoothly.

\item Generically, $\Gamma_1$ and $\Gamma_2$ must be mutually
nonlocal ($\langle \Gamma_1,\Gamma_2 \rangle \neq 0$) to have a
split flow (and hence a stationary BPS multicenter solution). A
degenerate exception occurs for mutually local charges when the
moduli at spatial infinity are already at a surface of marginal
stability: then the ``incoming'' branch of the split flow
vanishes, and a multicenter solution exists for arbitrary
positions of the centers.

\item Since the right-hand side of~(\ref{imZ1}) can only vanish for one
value of $\tau$, the phases $\alpha_i$ of the central charges
$Z_i$ will satisfy $|\alpha_1-\alpha_2|< \pi$ (at least if we put
$\alpha_1=\alpha_2$ at marginal stability, as opposed to $\alpha_1
= \alpha_2 + 2 n \pi$), even though separately, they do not have
to stay in the $(-\pi,\pi)$-interval.

\item This also implies that for mutually nonlocal charges, we can
rewrite (\ref{rmscond}) as
\begin{equation} \label{stabcond}
\langle \Gamma_1,\Gamma_2 \rangle \, (\alpha_1-\alpha_2) > 0 \,,
\end{equation}
where $\alpha_i = \arg Z(\Gamma_i)_{r=\infty}$. This is precisely
the stability condition for ``bound states'' of special lagrangian
3-cycles found in a purely geometrical setting by
Joyce~\cite{joyce} in the case where $\Gamma_1$ and $\Gamma_2$ are
special Lagrangian 3-spheres, and for values of the moduli
sufficiently close to marginal stability.

\end{itemize}

The above stability criterion is also quite similar to Douglas'
triangle stability criterion \cite{Dcat}, roughly as follows.
Consider three BPS charges $\bar{A}$, $B$ and $C$,\footnote{The
bar on A is for notational compatibility with \cite{Dcat}.} with
$C=\bar{A}+B$ and $\langle \bar{A},B \rangle
> 0$ (so $\langle B,C \rangle
< 0$ and $\langle C,\bar{A} \rangle < 0$). Identify Douglas'
``morphism grade'' between $P,Q \in \{A,B,C\}$ with $\phi_{PQ}
\equiv (\alpha_Q - \alpha_P)/\pi$, where $\alpha_P$ ($\alpha_Q$)
is the phase of $Z_P$ ($Z_Q$) and $\alpha_{\bar{A}} \equiv
\alpha_{A} + \pi$. Obviously, $\phi_{AB}+\phi_{BC}+\phi_{CA} = 1$.
By suitable labeling, we can assume $\langle \Gamma_1,\Gamma_2
\rangle >0$ for the composite state considered above, so we can
take $\bar{A}=\Gamma_1$, $B=\Gamma_2$ and $C=\Gamma$. The above
stability criterion for the composite state can now be rephrased
as $0<\phi_{AB},\phi_{BC},\phi_{CA}<1$, that is, in the
terminology of \cite{Dcat}, $(A,B,C)$ forms a ``stable triangle''.

Though there is an obvious similarity, this connection needs
further clarification. In particular, the role of morphism grades
outside the interval $(-1,1)$ is obscure in the supergravity
context at this point (see however section \ref{monprob}).

\subsection{Building the spectrum}

How can we determine whether there exists a (split) flow or not
for a given charge $\Gamma$ and vacuum moduli\footnote{With
``vacuum moduli'', we mean the moduli at spatial infinity.} $z_0$?

To find out if a single flow exists is no problem: basically, one
just has to check whether or not the central charge is zero at the
attractor point, as explained in section \ref{sec:single}. So one
can in principle determine algorithmically the single flow
spectrum at any point $z_0$ in moduli space.

The split flow spectrum is more difficult to obtain. Let us first
consider the simplest case, split flows with only one splitting
point, say $\Gamma \to \Gamma_1 + \Gamma_2$. At first sight, it
might seem that an infinite number of candidate constituents
($\Gamma_1$,$\Gamma_2$) has to be considered. However, the
situation is not that bad, at least if the mass spectrum of single
flows is discrete, does not have accumulation points, and is roughly
proportional to charge. This is what one would expect physically,
and we give an argument for this property of the spectrum in
appendix \ref{sec:strings}, based on an interesting link with the
multi-pronged string picture of quantum field theory BPS
states.\footnote{The finite region $E$ introduced in that appendix
consists here of a neighborhood of the single $\Gamma$-flow} Then
indeed, because $|Z(\Gamma)|$ is decreasing along the
$\Gamma$-flow, and at the split point we have
$|Z(\Gamma)|=|Z(\Gamma_1)|+|Z(\Gamma_2)|$, we only need to
consider pairs $(\Gamma_1,\Gamma_2)$ in the single flow spectrum
with mass less than $|Z(\Gamma)|_{\tau=0}$. Therefore, if the
single flow spectrum has indeed the above properties, we only need
to consider a finite number of cases.

In practice, to figure out precisely at which charge numbers one
can stop checking candidates, is of course a nontrivial problem on
its own. Nevertheless, in some cases it can be carried out without
too much difficulty, as we will illustrate in section
\ref{sec:spec}.

\subsection{Monodromies and multiple basins of attraction}
\label{sec:multibasin}

In \cite{M}, it was observed that a charge $\Gamma$ does not
necessarily have the same attractor point for all possible values
of the vacuum moduli: the moduli space (or more precisely its
covering space) can be divided in several different ``basins of
attraction''. Therefore, the corresponding black hole horizon area
and the BH-entropy are not just a function of the charge, but also
of the basin to which the vacuum moduli belong. This data was
called the ``area code'' in \cite{M}. This nonuniqueness might
seem somewhat puzzling, especially in the light of statistical
entropy calculations using D-branes. Also, what happens to the
supergravity solution when one moves from one basin to the other
seems rather obscure: do we get a catastrophe, a jump, a
discontinuity?

We will clarify these issues here, arriving once again at a
beautiful picture of how string theory resolves naive disasters.

\FIGURE[t]{\centerline{\epsfig{file=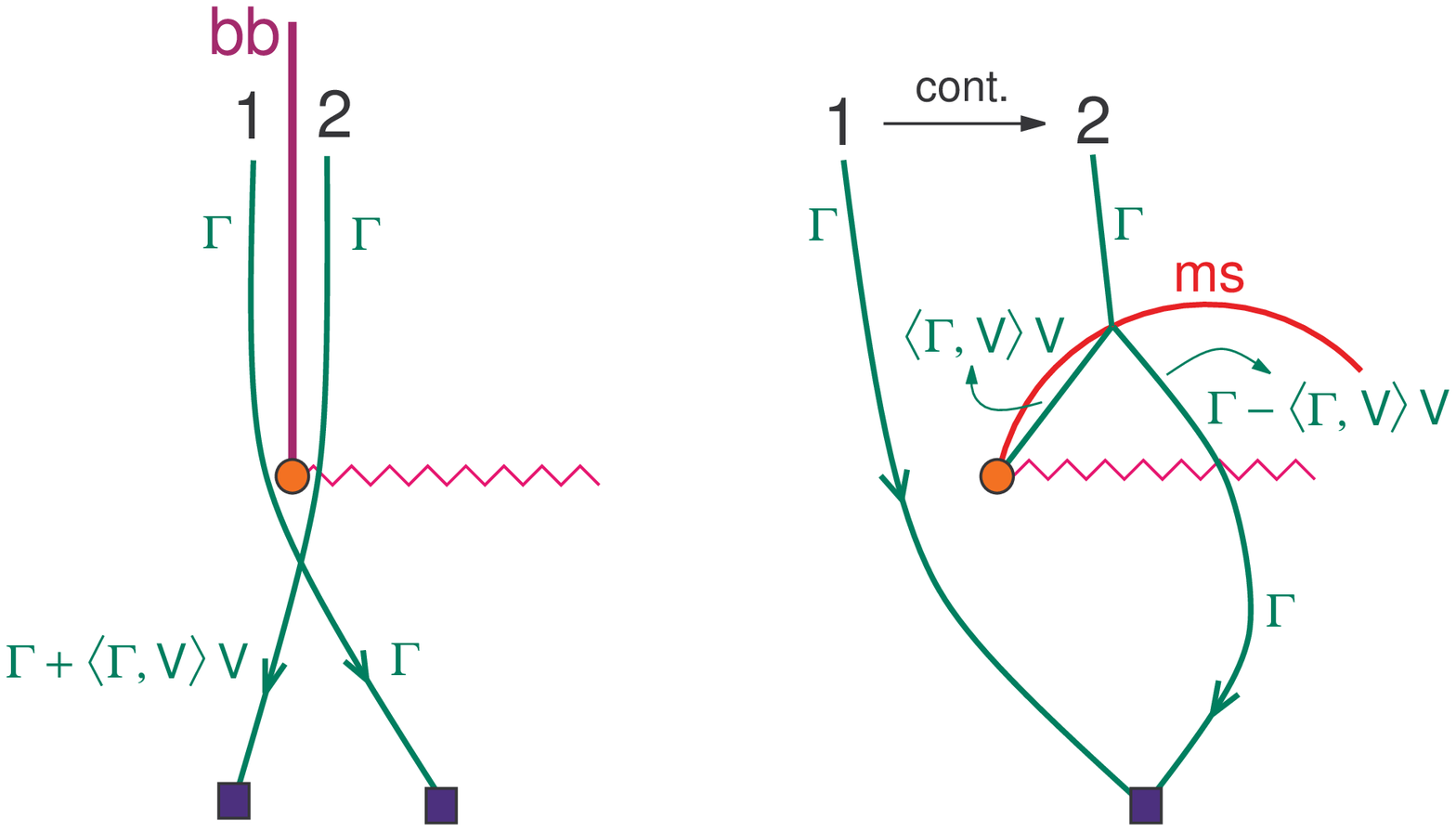,height=6cm}}
\caption{\emph{Left}: A conifold singularity (the red dot), with
vanishing cycle $V$, as source for different basins of attraction.
The flows to the left and the right of the basin boundary (labeled
``bb'') have different attractor points, due to the monodromy. The
horizontal wiggly line is a cut.  \emph{Right}: This is what
really happens when a flow is continuously moved from 1 to 2: it
transforms into a split flow with the new branch ending on the
singularity. The red line labeled ``ms'' is the corresponding
$(\langle \Gamma,V \rangle V,\Gamma-\langle \Gamma,V \rangle V)$-marginal stability line.}
\label{basinjump}}

The key observation is that jumps in the basin of attraction are
caused by the presence of singularities at finite distance in
moduli space, such as the conifold point for the quintic. Suppose
we have a singularity locus $z_s$ associated to a vanishing cycle
$V$, so that the monodromy about this singularity is given by
\begin{equation}
 \Gamma \to \Gamma+ \langle \Gamma,V \rangle \, V \, .
\end{equation}
Consider, as shown in fig.\ \ref{basinjump}, a $\Gamma$-flow that
passes just to the left of $z_s$, with $\langle \Gamma,V \rangle
\neq 0$. Now move the starting point of the flow to the right, as
if we were trying to ``pull'' the flow through the singularity. We
will not succeed to do this smoothly, because of the monodromy: a
flow starting off with charge $\Gamma$ and passing just to the
right of $z_s$ can no longer be assigned charge $\Gamma$ at points
beyond the singularity; instead, we should assign it charge
$\Gamma'= \Gamma+ \langle \Gamma,V \rangle \, V$. One way to
understand this is that the second flow necessarily crosses the
cut starting at $z_s$ (we assumed implicitly that the first one
does not; this is of course purely conventional). This means that
this flow will no longer converge to a $\Gamma$-attractor point,
but rather to a $\Gamma^\prime$-attractor point, which in general
will be different from the original one, with a different value
for $|Z|_{\mathrm{min}}$, possibly even a regular zero (in that
case, no single center BPS solution exists anymore). Thus we get
two basins of attraction, with boundary formed by the ``critical''
flow, i.e. the $\Gamma$-flow hitting the singularity $z_s$.

Does this mean the solution and its intrinsic properties such as
entropy jump discontinuously when we vary the vacuum moduli in
this way, possibly even kicking the state out of the spectrum
(without obvious decay products)? The answer is no. As explained
in more detail in \cite{branessugra}, what really happens is that,
upon moving the flow through the singularity, $n=\langle V,\Gamma
\rangle$ units of $V$-charge are created at the locus in space
where the moduli acquire the singular value $z_s$.\footnote{This
is somewhat similar to the creation of charge at the center of a
dynamically occurring flop transition in $M$ Theory, necessitated
by the presence of four-form flux, as recently studied in
\cite{Mflops}.}

This is consistent with energy conservation, because at this
locus, $V$ particles are massles. It is also consistent with
charge conservation, because of the subtleties associated to the
monodromy.  In the case of the example at hand, when one continues
varying the vacuum moduli, the newly born $V$-particles will
acquire mass, and a full-fledged multicenter solution (of the type
described in the previous section) emerges. The black hole core
remains unchanged however, and all fields change smoothly during
the transition.

In the split flow picture (see also fig.\ \ref{basinjump}), what
happens is that the attractor flow gets a new branch, of charge $n
V$, terminating on the singularity locus (corresponding to a
(multi) empty hole constituent). Note that because $Z(V)$ is zero
at $z_s$, there will always be a surface (or line) of $(V,\Gamma)$
marginal stability starting at $z_s$, as is needed for the split
flow to exist. Furthermore, if one tried to continue circling
around the singularity, one would unavoidably cross this surface
of marginal stability, and a decay would result. Thus, mondromies
about singularities of this kind will always induce decay of the
configuration (if one is sufficiently near the singularity). This
does not mean that this charges disappears from the spectrum:
other BPS configurations with this charge can still exist.

Similar things could happen near different kinds of singularities
(not of ``conifold''-type), though not necessarily so. For
instance, circling around the large complex structure point of the
quintic with a generic flow will not do anything spectacular,
essentially because the LCS point is at infinite distance in the
moduli space, making it impossible for the flow to ``cross''.
Instead, it will just get wrapped around it.

The phenomena described in this section are completely analogous
to what happens in the transition from simple to three-pronged
strings in the description of QFT BPS states
\cite{BPS37,threeprong}, though it arises here from a quite
different starting point.\footnote{In the very recent work
\cite{argyres}, the connection with this picture at the effective
field theory level was investigated in detail.} The effective
string action of appendix \ref{sec:strings} makes this analogy
quite precise, allowing us to carry over many of the insights
obtained in that context.

\section{Type IIA string theory on the Quintic and its IIB mirror}\label{sec:quintic}
\setcounter{equation}{0}

In the remainder of this paper, we will apply the above general
considerations to type IIA string theory compactified on the
quintic Calabi-Yau $X$, or equivalently type IIB compactified on
its mirror $Y$.

The quintic has $h^{2,1}(X)=101$ and $h^{1,1}(X)=1$. Consequently,
the four dimensional low energy supergravity theory of a IIA
compactification on $X$ has one vector multiplet, where the
complex scalar corresponds to the complexified K\"ahler modulus of
$X$. Its mirror manifold $Y$ has $h^{2,1}(Y)=1$ and
$h^{1,1}(Y)=101$, and the complex scalar of the vector multiplet
in the IIB low energy theory corresponds to the complex structure
modulus of $Y$. The manifold $Y$ is defined by a single
homogeneous equation of Fermat type:
 \be \label{quinticdef}
  {\sum_{i=1}}^{5}{x_{i}^5}-5 \, \psi \, {x_{1}x_{2}x_{3}x_{4}x_{5}}=0
 \ee with $x_{i}$ homogeneous coordinates on $\C P^4$ and a single
complex parameter $\psi$, the complex structure modulus. More
precisely, $Y$ is the quotient of this algebraic variety by the
identifications $x_i \simeq \omega^{k_i} x_i$, with $\omega=e^{2
\pi i/5}$ and the $k_i \in \IZ$ satisfying $\sum_i k_i = 0$. Note
also that the $\psi$-plane is actually a 5-fold covering of moduli
space, since $\psi$ and $\omega \psi$ yield isomorphic spaces
through the isomorphism $x_1 \to \omega x_1$.

\subsection{Quantum Volumes and Meijer Functions}

For the sake of future generalizations, we will start by
formulating our analysis of the quintic in the rather general
framework of \cite{GLaz} and the work upon which it is drawn.
Readers desiring a more explicit treatment of these matters are
encouraged to consult the reference given above.

Define the \emph{quantum volume} \cite{quantvol} of a holomorphic 
even dimensional
cycle in an algebraic variety with trivial anticanonical bundle to
be equal to the quantum corrected mass of the (IIA) BPS saturated
D-brane state wrapping it; this is equal to the \emph{classical}
mass of its mirror 3-cycle.  In the large radius limit, this
prescription agrees with our naive notion of volume, but as we
move into the quantum regime, corrections arise which severely
alter the behavior of these volumes as functions of the moduli,
away from what one would expect.  In this manner, we may obtain a
quantum mechanically exact expression for the volume of a given
even dimensional holomorphic cycle $\gamma$ in a variety $X$ with
trivial anti-canonical bundle, in terms of the normalized period
(\ref{Zdef}) of its mirror 3-cycle $\Gamma$:
 \be
  \label{BPSmassesinIIB} V(\gamma) = M(\gamma)= |Z(\Gamma)| =
  \frac{|\int_{\Gamma}{\Omega(z)}|}{(\int_{Y}{i \, \Omega(z)\wedge
  {\overline \Omega(z)}} )^{1/2}}~~=
  \frac{|q^i\int_{\Gamma_i}{\Omega}|}{(\int_{Y}{i \, \Omega(z)\wedge
  {\overline \Omega(z)}})^{1/2}} ~~.
 \ee
In the above, $\{\Gamma_i \}_i$, $i=1,\ldots,2 h^{2,1}(Y) + 2$, is
an integral basis of $H_{3}(Y)$, $\Omega (z)$ is the holomorphic
three-form, written with explicit dependence on the moduli $z$,
$q^i$ are the integral charges of the cycle with respect to the
$\Gamma_i$, and $\int_{\Gamma_i}{\Omega(z)}$ are the periods of
the holomorphic three-form.

In a model where $h^{1,1}(X)=1$, such as the quintic, there is
only one modulus, and we may identify a point $z=0$ with the so
called large complex structure limit, i.e. the complex structure
for the mirror variety $Y$ which is mirror to the large volume
limit of $X$.  In this class of examples, holomorphic cycles of
real dimension $2j$ on $X$ are mirror to three-cycles on $Y$ whose
periods have leading ${\rm log}^j z$ behavior near $z = 0$. Thus,
finding a complete set of periods of $\Omega(z)$ and classifying
their leading logarithmic behavior gives us a means of identifying
the dimension of their even-cycle counterpart on $X$.  In this
context, when we speak about a cycle on $X$ of real dimension
$2j$, we refer to a cycle on $X$ with $2j$ being the maximal
dimensional component, but with the identity of the various lower
dimensional ``dissolved'' cycles left unspecified. More input is
needed to identify the latter.

The technology of Meijer periods \cite{Meijer_refs, Norlund}
allows us to write down a basis of solutions to the Picard-Fuchs
equation associated to a given variety $Y$, which is indexed by
the leading logarithmic behavior of each solution.  In particular,
we are able to find a basis of solutions, each representing a
single BPS $2j$ brane on $X$, viewed in terms of the mirror
variety $Y$.  These periods will have branch cut discontinuities
on the moduli space; only on the full Teichm\"{u}ller space they
are continuous.

The periods of the holomorphic three-form on a Calabi-Yau manifold
are solutions to the generalized hypergeometric equation:

\be
\label{HG_eq} \left[ \delta~\prod_{i=1..q}{(\delta+\beta_i-1)}-
z~\prod_{j=1..p}{(\delta+\alpha_j)} \right]u=0~~. \ee

\noindent where $\delta=z\partial_{z}$ and $\alpha_{i},\beta_{j}$
are model dependent constants.

For a given non linear sigma model in the class of varieties which
are algebraic Calabi-Yau complete intersections, one may easily
read off the form of the hypergeometric function having regular
behavior under monodromy, and from that determine the form of the
hypergeometric equation the periods satisfy.

\subsection{Periods, monodromies and intersection form for the quintic}
\label{sec:quinticperiods}

The model dependent parameters in (\ref{HG_eq}) for the (mirror)
quintic (\ref{quinticdef}) are $\alpha=\{1/5,2/5,3/5,4/5\}$ and
$\beta=\{1,1,1\}$, and $\psi$ is related to $z$ by $z=\psi^{-5}$.
A class of solutions to these PDE manifest themselves as Meijer
functions $U_j$, each with ${\rm log}^j z$ behavior around $z =
0$. For the quintic, they have the following integral
representation:

\be\label{intrep} \label{UJ} U_j(z)=\frac{1}{(2\pi
i)^j}\oint_{\gamma}\frac{ds
\Gamma(-s)^{j-1}\prod_{i=1}^{4}\Gamma(s+\alpha_i) ((-1)^{j+1}z)^s~
}{\Gamma(s+1)^{3-j}}, \ee

\noindent for $j\in\{0,1,2,3 \}$, with $\alpha$ as defined above.

The integral above has poles at $\alpha_{i} -s = -n$ and
$\beta_{j}+s = -n$ for $n\in \IZ^+$.  We may evaluate it by the
method of residues by choosing $\gamma$, a simple closed curve,
running from $-i\infty$ to $+i\infty$ in a path that separates the
two types of poles from one another.  Closing the contour $\gamma$
to the left or to the right will provide an asymptotic expansion
of $U_{j}(z)$ which is adapted to either the Gepner point
($\psi^{-5}=z=\infty$ in this parametrization) or the large
complex structure point ($z=0$).  Our choice of defining
polynomial for the mirror quintic is such that the discriminant
locus (conifold point) lies at $z=1$. The detailed expression of
the periods $U_j(z)$ in terms of the predefined Meijer functions
in \emph{Mathematica} can be found in appendix \ref{numerics}.

For the quintic, using the conventions detailed in \cite{GLaz} we
have monodromy matrices around these regular singular points,
given in the basis of (\ref{intrep}) by:

{\footnotesize \bea
\begin{array}{ccc}
T[0]=\left [\begin {array}{cccc} 1&0&0&0\\-1 &1&0&0\\1&-1 &1&0
\\0&0&-1 &1\end {array}\right ]&~,~&
T[\infty]=\left [\begin {array}{cccc} -4&5&-5&5\\ -1&1&0&0\\1&-1
&1&0\\ 0&0&-1 &1\end {array} \right ]\end{array}\nn \eea}

\noindent and $T[1]=T[\infty] \cdot T[0]^{-1}$ for $\im z < 0$,
$T[1]=T[0]^{-1} \cdot T[\infty]$ for $\im z > 0$.

We use the conventions of  \cite{D} to assign precise D-brane
charges to a given state.  To that end, we will work in a basis
where we label the charge of a state as $(D6,D4,D2,D0)$, i.e.\
$q=(q_6,q_4,q_2,q_0)$. We will call the corresponding period basis
$\Pi$. The elements of this basis are related to the Meijer basis
$U$ by $\Pi=L\cdot U$ where $L$ is the following matrix:

{\footnotesize \bea L={\frac{8 \,i \pi^3}{125}} \left [\begin
{array}{cccc} 0&5&0&5\\ 0&1&-5&0\\ 0&-1 &0&0\\ 1&0&0
&0\end{array}\right ] \eea }

\noindent The cycles corresponding to the periods $\Pi$ have the
following intersection form $Q$:
 {\footnotesize \bea Q = \left [\begin
 {array}{cccc} 0&0&0&-1\\ 0&0&1&0\\ 0&-1 &0 &0\\ 1&0&0&0\end
 {array} \right ].
 \eea}
The monodromies around the large complex structure, Gepner and
conifold points (the latter for $\im z < 0$) in this basis are:
 {\footnotesize \bea
  \begin{array}{ccccc}
  T[0]=\left [\begin{array}{cccc} 1&1&3&-5\\0&1&-5&-8\\0&0&1&1
  \\0&0&0&1\end{array}\right ]&~,~&
  T[\infty]=\left [\begin{array}{cccc} 1&1&3&-5\\ 0&1&-5&-8\\0&0
  &1&1\\ 1&1&3&-4\end{array} \right ]&~,~&
  T[1]=\left [\begin{array}{cccc} 1&0&0&0\\ 0&1&0&0\\0&0
  &1&0\\ 1&0&0&0\end{array} \right ]
  \end{array}\nn
 \eea}
\noindent From the form of the $T[1]$-monodromy, it follows that
the BPS state becoming massless at the conifold point gets
assigned charge $(D6,D4,D2,D0)$ = $(1,0,0,0)$ at $z=e^{i 0 -}$
($\psi = e^{i 0 +}$). Similarly, from $T[1]$ for $\im z > 0$, one
can deduce that this BPS state gets assigned charge $(1,1,3,-5)$
at $z=e^{i 0 +}$ ($\psi = e^{2 i \pi/5 \, \, -}$). The charge
ambiguity is mathematically due to the choice of cuts, and
physically due to the fact that only at large radius in the type
IIA theory is the geometric labeling of D-brane charges really
meaningful.

We may use the above to calculate the K\"{a}hler potential
(\ref{kahlerpotential}), so as to obtain the periods with respect
to the normalized holomorphic three-form $\Omn$:

\be
 e^{-\CK}=i \, {\Pi(z)}^{\dagger} \cdot {Q}^{-1} \cdot \Pi(z) \, .
\ee

\noindent Then the correctly normalized central charge $Z(q)$ for
a a charge $q = (q_6,q_4,q_2,q_0)$ is
 \be
Z(q)(z,\bar{z})=e^{{\CK(z,\bar{z})}/2} \, q \cdot \Pi(z)
 \ee
\noindent Note that this normalization destroys the holomorphicity
of the periods in question, allowing them to possess local minima
of their norm that are nonzero.

\section{Practical methods for computing attractor flows}\label{sec:computation}
\setcounter{equation}{0}

Setting up an efficient scheme to compute attractor flows is of
course of prime importance in numerical studies. We will explain
the essential features of our strategy here, but the reader who is
only interested in the final results can skip this section.

We have followed two complementary approaches, essentially based
on the two different forms of the attractor flow equations.

The first form, equations (\ref{at1})-(\ref{at2}), suggests (at
least for one-parameter models) to compute the flows using a
step-by-step steepest descent method, which is a refinement of
brute minimization of the absolute value of the central charge.
This refinement is needed because the flow can cross one or more
cuts in the moduli space, so care has to be taken that the correct
minimizing path is followed, especially in the light of the
existence of several basins of attraction.

The second form, equation (\ref{integrated}), gives an algebraic
way to compute the flow. This is somewhat more involved, but has
the advantage that no accumulation of numerical errors occurs. In
practice, this also means that this method is faster, basically
because the path can be computed in larger chunks. It is also easy
to compute the precise space-dependence of the metric and moduli
fields using this method, and it is in principle straightforward
to generalize it to higher dimensional moduli spaces. The steepest
descent method on the other hand has the advantage that no
equations have to be solved numerically (only step-by-step
minimization is needed), making the procedure somewhat more
robust, as sometimes the algorithm one uses to compute the zeros
of an equation fails to converge.

\subsection{Method of Steepest Descent} \label{sec:descent}

For one-parameter models, the attractor equations
(\ref{at1},\ref{at2}), simplify greatly.  Since the metric on the
moduli space has only a single component $g_{z\bar{z}}$,
(\ref{at2}) is reduced to:
 \be\label{simpleat}
 \partial_{\tau} z=-\rho \, \partial_{\bar{z}}|Z(q)|
 \ee
where $\rho > 0$ denotes the metric factor.

The metric factor $\rho$ will only change the speed at which an
integral curve of this equation is traversed, not the path itself,
so the $\rho$ dependence may be undone by a reparametrization of
the ``proper time'' parameter $\tau$.  Thus, the attractor flow
lines will be exactly the lines of steepest (flat) gradient
descent in moduli space.\footnote{Notice that in a model with
$h^{1,1}>1$, we will have a greater number of metric components,
which cannot be in general be ignored.}

\FIGURE[t]{\centerline{\epsfig{file=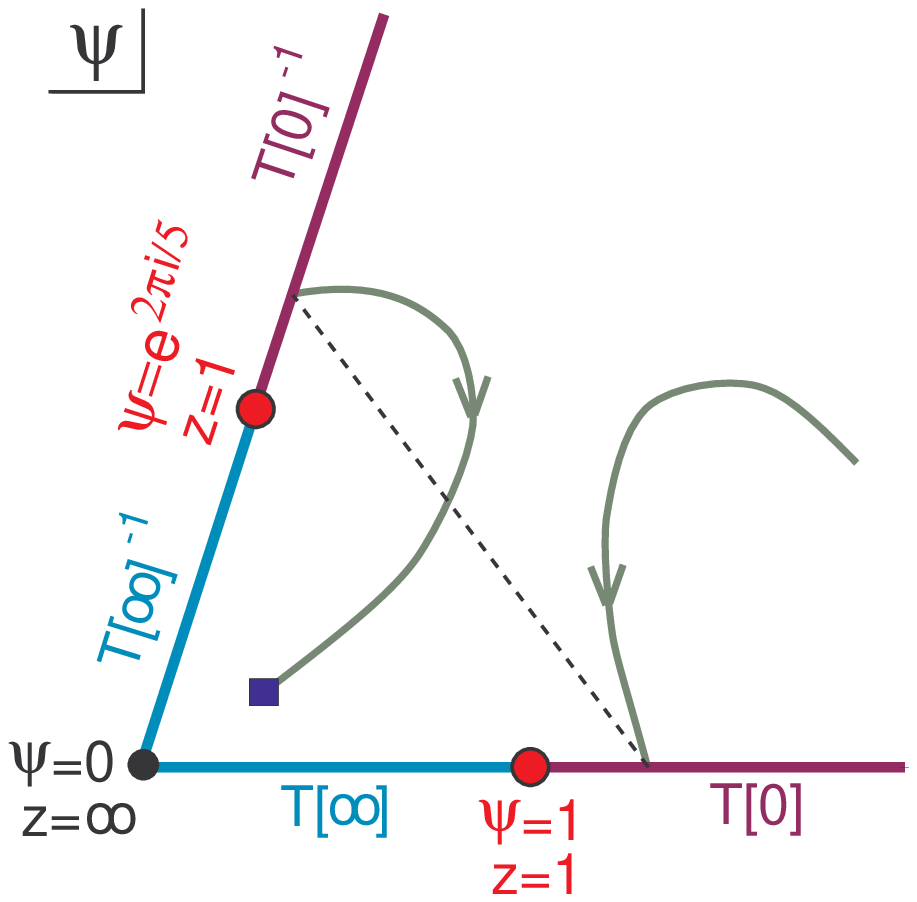,height=7cm}}
\caption{Sketch of the ``fundamental wedge'' $0 < \arg \psi < 2
\pi/5$ in the $\psi$-plane, a section of Teichm\"{u}ller space
unambiguously parametrized by $z=\psi^{-5}$. Attractor flows have
definite charge and are continuous on Teichm\"uller space, but if
we want to represent them exclusively on the fundamental domain,
the assigned charge jumps at the wedge boundaries as $q \to q.M$,
with $M$ the appropriate monodromy matrix. The labels $T[0]$ and
$T[\infty]$ along each branch cut in the picture indicate the
monodromy matrix to be applied for $|z|<1$ resp.\ $|z|>1$.
Indicated is an attractor flow line which undergoes a monodromy
from a charge $q$ to a charge $q \cdot T[0]$.}\label{section}}

This observation can be used to compute numerically the attractor
path in moduli space. Suppose we begin with a charge $q$, at an
initial modulus $\psi=\psi_0$ in the fundamental domain of the
$\psi$-plane (see fig.\ \ref{section}). The next point in the path
is then approximately given by the point with the smallest
$|Z(q)|$ on a small circle around the initial point. By repeating
this for a circle around this new point, we find the second point,
and so on, producing the approximate path of steepest descent.
When a branch cut is approached, at $\arg \psi\in\{0, \frac{2\pi}{5}\}$,
we must transform $q$ by an appropriate monodromy as we pass
through the cut.  If we are traveling down through a cut and
$|z|<1$, $T[0]$ should be applied (by right multiplication)
to $q$ to determine the new charge, while if $|z|>1$, we should
apply $T[\infty]$, as shown in fig.\ \ref{section}. Traveling up
through a cut requires us to apply the inverses of these matrices,
respectively. The procedure stops when a local minimum of $|Z(q)|$
is reached, that is, at the attractor point.

Notice that it is important to use this step-by-step minimization
procedure rather than an arbitrary minimization algorithm, since,
as explained in section \ref{sec:multibasin}, the presence of the
conifold point can induce distinct basins of attraction. Therefore
particular care should also be taken in the numerical procedure
when the flow comes near a conifold point:\footnote{Incidentally,
in many examples, the conifold point tends to squeeze the
attractor flows in its neighborhood towards it, making the cases
where the flow comes very closely to this point far from
non-generic.} a too low resolution of the path steps can result in
the wrong monodromy matrix being applied to the charge, yielding
an incorrect final result for the attractor point.

\subsection{Using the integrated BPS equations}

The second method is based on equation (\ref{integrated}). Suppose
we want to compute the flow for a charge $q \equiv
(q_6,q_4,q_2,q_0)$, starting from $z=z_0$ (or $\psi=\psi_0$).
Define for an arbitrary charge $q'$ the real function $f(q',z)$ as
\begin{equation}
 f(q',z) = \im [Z(q') \overline{Z(q)}] \, .
\end{equation}
Let $\ell_1$ and $\ell_2$ be two charges that are mutually local
with respect to $q$ (i.e.\ $\langle \ell_i,q \rangle=0$), and form
together with $q$ a linearly independent set. If $q_6$ and $q_4$
are not both zero, we can take for instance
\begin{eqnarray}
 \ell_1 &=& (q_4,-q_6,-q_0,q_2) \label{L1} \\
 \ell_2 &=& (0,0,q_6,q_4) \,. \label{L2}
\end{eqnarray}
If $q_4=q_6=0$, we can take $\ell_2 = (q_1,q_2,0,0)$ instead.
Finally, let $d$ be a (not necessarily integral) charge dual to
$q$, i.e.\ such that $\langle d,q \rangle = 1$. For instance, we
can take
\begin{equation} \label{dualcharge}
 d=(-q_0,q_2,-q_4,q_6)/(q_0^2+q_2^2+q_4^2+q_6^2) \, .
\end{equation}
A convenient parameter for the flow turns out to be
\begin{equation} \label{mudef}
 \mu \equiv e^U \frac{|Z(q)|}{|Z(q)|_0} \, ,
\end{equation}
where $|Z(q)|_0 \equiv |Z(q)|_{z=z0} = |Z(q)|_{\tau=0}$. This
parameter always runs from $1$ to $0$, no matter what nature of
the attractor point is (zero central charge or not).

Taking the intersection product of (\ref{integrated}) with the
$\ell_i$ gives, after some reshuffling:
\begin{equation}
 f(\ell_i,z) = f(\ell_i,z_0) \, \mu \, .
\end{equation}
This is a system of two equations, which can easily be solved
(numerically) for $z$ as a function of $\mu$,\footnote{In
practice, numerically solving this system requires two starting
points $z_1$, $z_2$. The algorithm then tries to find a root near
$z_1$, $z_2$. To guarantee that the (right) root is found for a
given value of $\mu$ not close to $1$, it is necessary to solve
this system in several steps, starting at $\mu=1$ and gradually
lowering $\mu$ down to the desired value, taking the roots last
found as the new starting points. Following this procedure down to
$\mu=0$, one walks around in moduli space, possibly crossing
several cuts, till one finally arrives at the attractor point of
the attractor basin under consideration. The same cautionary
remarks as in section \ref{sec:descent} apply near a conifold
point.} yielding the desired attractor flow $z(\mu)$ in moduli
space. The value of the metric factor $U(\mu)$ along the flow can
be computed directly from (\ref{mudef}). Finally, to get the
$\tau$-dependence, take the intersection product of
(\ref{integrated}) with the dual charge $d$. This yields:
\begin{equation}
 \tau(\mu) = 2 [f(d,z_0)-f(d,z(\mu))/\mu] \, .
\end{equation}
Thus we obtain the complete solution to the attractor flow
equations.

\section{Analysis of the quintic}\label{sec:spec}
\setcounter{equation}{0}

\subsection{Some notation and conventions} \label{sec:notconv}

We will usually work on the $\psi$-plane or the $w$-plane (see
below) to describe attractor flows. We take the wedge $0<\arg \psi
< 2 \pi/5$ to be the fundamental domain. Charges will be given in
the IIA $(D6,D4,D2,D0)$ $\Pi$-basis of section
\ref{sec:quinticperiods}. Often however, especially close to the
Gepner point $\psi=0$, it is more transparent to give a label
based on the $\IZ_5$ monodromy around the Gepner point. In
general, we will use the notation $(q_6,q_4,q_2,q_0)^n$ for the
charge $(q_6,q_4,q_2,q_0).T[\infty]^n$, with $n \in \IZ
\mathrm{~mod~} 5$. For example $(1,0,0,0)^1=(1,1,3,-5)$, which is
the state becoming massless at $\psi=e^{2 i \pi/5}$. More
generally, $(1,0,0,0)^n$ becomes massless at $\psi=e^{2 n i
\pi/5}$. In the type IIB picture, these states correspond to
3-branes wrapped around the appropriate vanishing conifold cycle.

\FIGURE[t]{\centerline{\epsfig{file=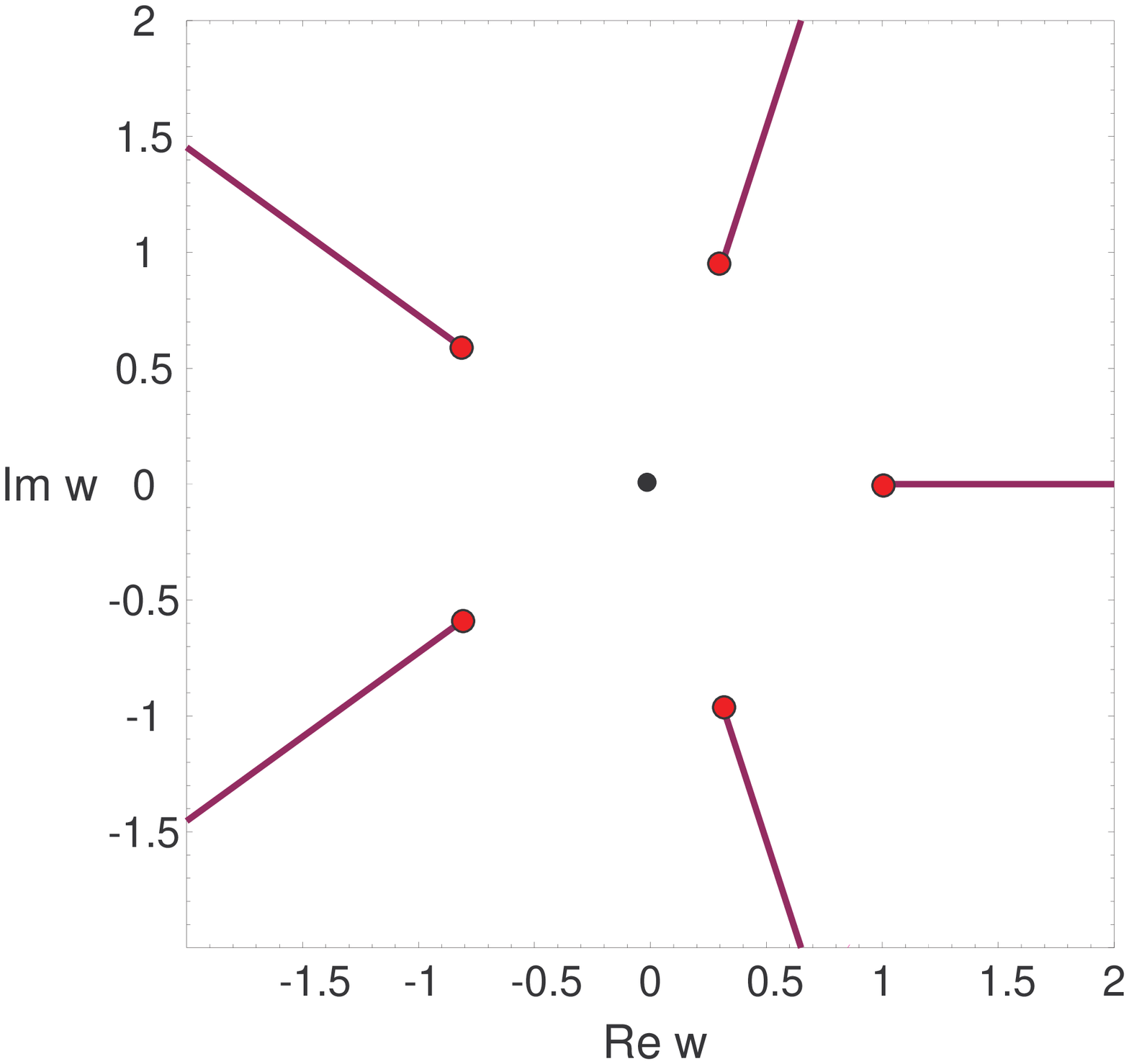,height=6.5cm}}
\caption{The $w$-plane is a 5-fold cover of moduli space, and is
especially convenient for plotting flows. The five fat purple
lines are period cuts starting at the conifold point copies
$w=e^{2 n i \pi/5}$ and running to the large complex structure
limit. The black dot in the middle is the Gepner point $\psi=w=0$.
Since all periods can be taken to be continuous around $w=0$, we
only indicate cuts running to $w=\infty$.}\label{wplane}}

For graphing purposes, we find it convenient to work with a
non-holomorphic coordinate $w$ on moduli space, defined as
\begin{equation}
 w \equiv \frac{\ln(|\psi|+1)}{\ln 2} \, \frac{\psi}{|\psi|} \, .
\end{equation}
This coordinate is proportional to $\psi$ close to the Gepner
point, and grows as $\ln |\psi|$ in the large complex structure
limit, so we get essentially power-like dependence of the periods
on $w$ in both regimes. The normalization of $w$ is chosen such
that the copies of the conifold point are located at $w=e^{2 n i
\pi/5}$, $n=0,\ldots,4$ (see fig.\ \ref{wplane}).

We will freely mix IIA and IIB language. For instance, we refer to
the limit $\psi \to \infty$ both as the large complex structure
limit (IIB) and as the large radius limit (IIA).

Our conventional path for interpolating between $\psi=0$ and
$\psi=\infty$ is the line $\psi = e^{i \pi/5} \IR^+$.

We will mostly focus on the analysis of states with low charges,
and use supergravity language to describe the corresponding
solutions, though the supergravity approximation cannot
necessarily be trusted in those cases. However, they can always be
trivially converted to large charge solutions by multiplying the
charge with a large number $N$, and correspondingly scaling all
lengths with a factor $N$. We will come back to this in the
discussion section.

\subsection{Single flow spectra} \label{sec:singleflowspectra}

\FIGURE[t]{\centerline{\epsfig{file=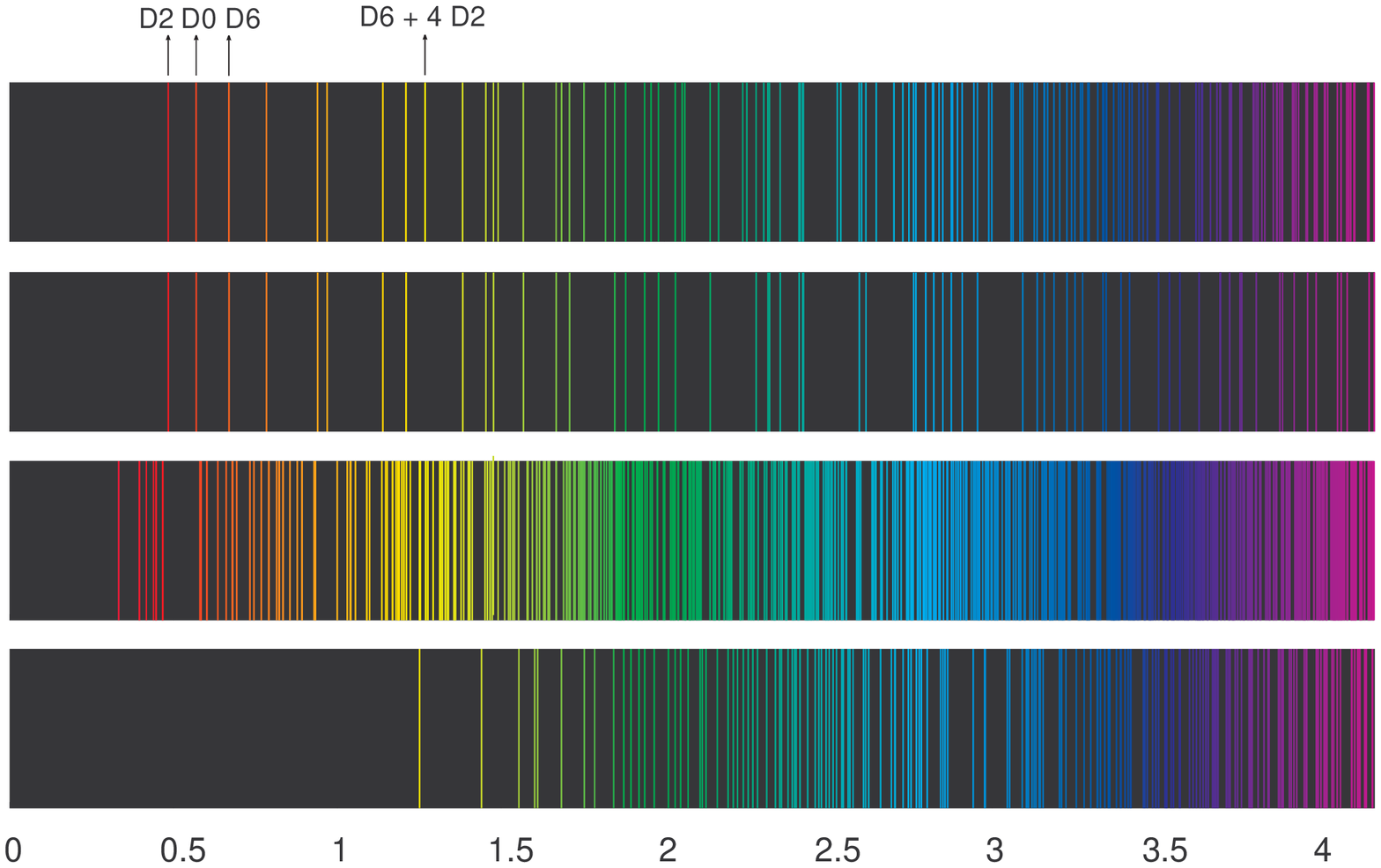,height=10cm}}
\caption{From top to bottom: {\bf (1)}: spectrum of the lightest
single flows originating at $\psi=0$. The horizontal axis
gives the mass in four dimensional Planck units. The state with
pure $D2$ charge (and its $\IZ_5$ images) has the lowest mass,
followed by the $D0$, the $D6$ (which is an empty hole), and five
mixed $(D2,D0)$ states. The lightest black hole comes next, with
charge $(1,0,4,0)$. {\bf (2)}: Same as (1), excluding the regular
black holes. {\bf (3)}: Like (1), but starting at the point $\psi =
0.0851-0.3997 i$, i.e.\ the ``crash point'' of fig.\
\ref{MrDecay}. Here the state $(1,0,0,0)^{-1}$ is the lightest.
Note that the spectrum lines are more ``spread out'' here than in
(1), due to the fact that the $\IZ_5$-symmetry is broken, causing
the lines to split in five. {\bf (4)}: Masses of regular BPS black
holes that exist at the Gepner point, evaluated \emph{at} their
attractor points, where they acquire their minimal value. The
lightest charge here is again $(1,0,4,0)$, with $M_{min} \approx
1.250947$. }\label{spectrumlines}}

A first step in analyzing the spectrum from the attractor flow
point of view is determining which charges give rise to a
well-behaved single flow for a given vacuum modulus, that is,
flows not crashing on a regular zero. In other words, these are
charges that have a regular BPS black hole solution, or a BPS
empty hole solution, or a BPS solution with a mild point-like
naked singularity. The latter correspond to $(n D2,m D0)$-charges,
which have their attractor point at large radius $\psi=\infty$
\cite{M}.  Though the four dimensional supergravity approximation
breaks down close to their center (as the quintic decompactifies
there), we will consider these solutions to be admissible and in
the physical BPS spectrum.
%
Note that the central charges of these $D2-D0$ particles vanish in
the large radius limit, so they become massless in four
dimensional Planck units. However, since the internal space
decompactifies in this limit, the natural scale is the \emph{ten}
dimensional Planck mass, with respect to which the mass of these
particles stays finite (for a pure $D0$) or diverges (if $D2$
charge is involved).

A single flow spectrum analysis is illustrated in fig.\
\ref{spectrumlines}, which resulted from a scan of all charges
$(q_6,q_4,q_2,q_0)^n$, with the $q_i$ between $-5$ and $5$, and
$n$ between $0$ and $4$, together a set of $71,070$ charges (which
do not have to be considered all separately, as there are obvious
redundancies, such as inversion of all charges and the
$\IZ_5$-symmetry at the Gepner point). Of this set, $31040$
charges can be realized as a single flow.

\FIGURE[t]{\centerline{\epsfig{file=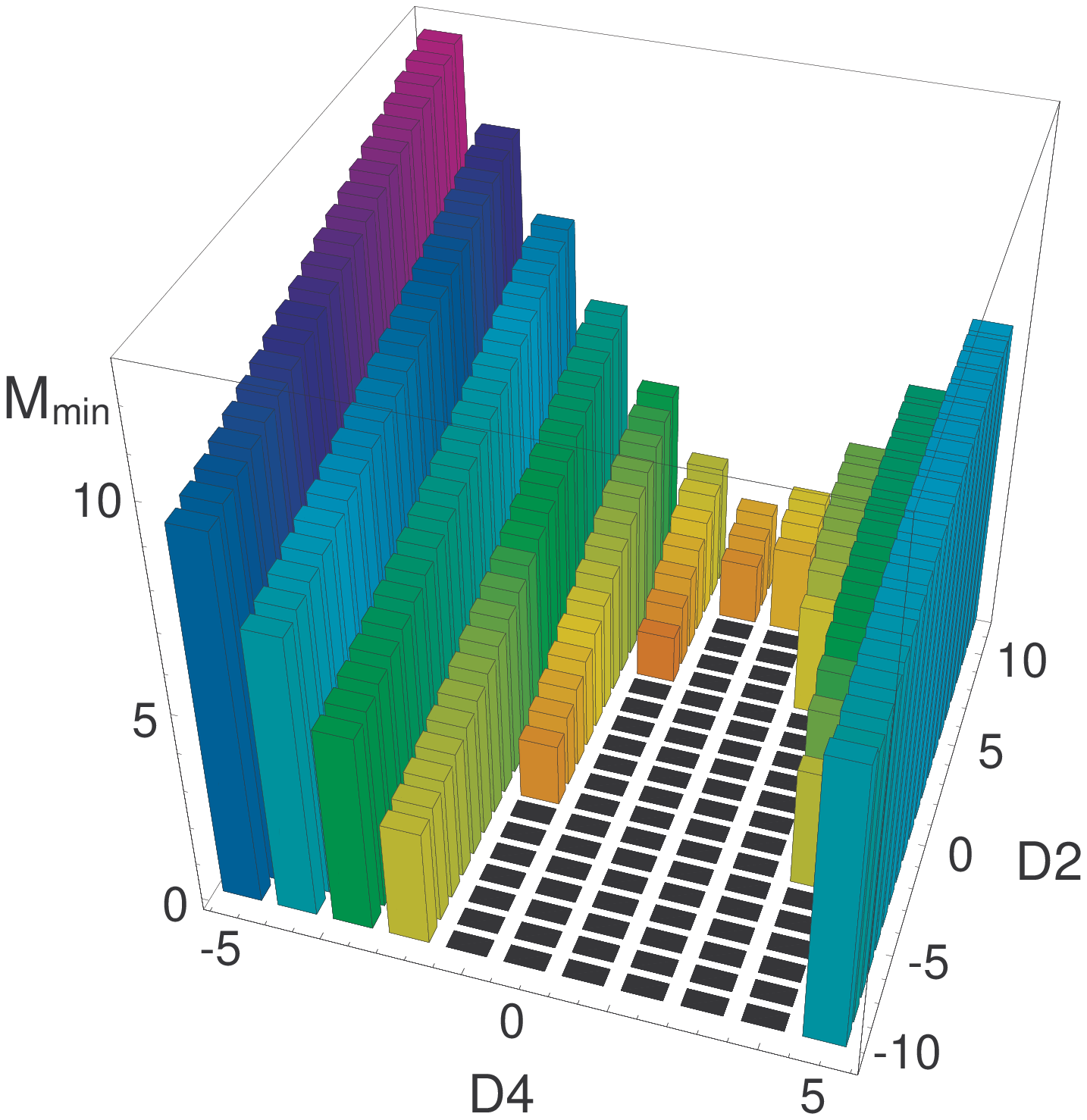,height=7cm}}
\caption{Masses $M_{min}$ evaluated at the attractor points for a
set of charges $(1,q_4,q_2,0)$, with flows starting at $\psi=0$. A
black bar of zero height indicates that the flow does not
exist.}\label{barspec}}

As expected from the general arguments in appendix
\ref{sec:strings}, our numerical data indicates that the masses of
the BPS states indeed tend to grow with charge (see also fig.\
\ref{barspec}). This is not trivial, since it is \emph{not} true
for the mass of arbitrary \emph{candidate} BPS charges. At the
Gepner point, the BPS mass of a charge $q$ is given by
\begin{eqnarray}
 M(q)= c \, \left| (-q_4+q_6) \, \omega^{-1} + (-q_0+q_2-5q_4-2q_6) +
 (q_0+2q_4+q_6) \, \omega + q_4 \, \omega^2 \right| \, , \nonumber
\end{eqnarray}
with $c=[ 5(5+\sqrt{5})/2 ]^{-1/4}$ and $\omega=e^{2 i \pi/5}$.
The different terms in this expression correspond to the
components of $q$ with respect to the basis $\{ (0,0,1,0)^n \}$,
$n=-1,\ldots,2$. Note that in particular
\begin{equation}
 M({q_6,0,q_2,0})=c \, |q_2 -\frac{5-\sqrt{5}}{2} q_6| \, ,
\end{equation}
so no such BPS states should exist at the Gepner point with
$q_2/q_6$ too close to $\frac{5-\sqrt{5}}{2} \approx 1.38197$.
This is consistent with our single flow results, as illustrated
fig.\ \ref{barspec}.

\subsection{Example of black hole decay from LCS to Gepner point}

\FIGURE[t]{\centerline{\epsfig{file=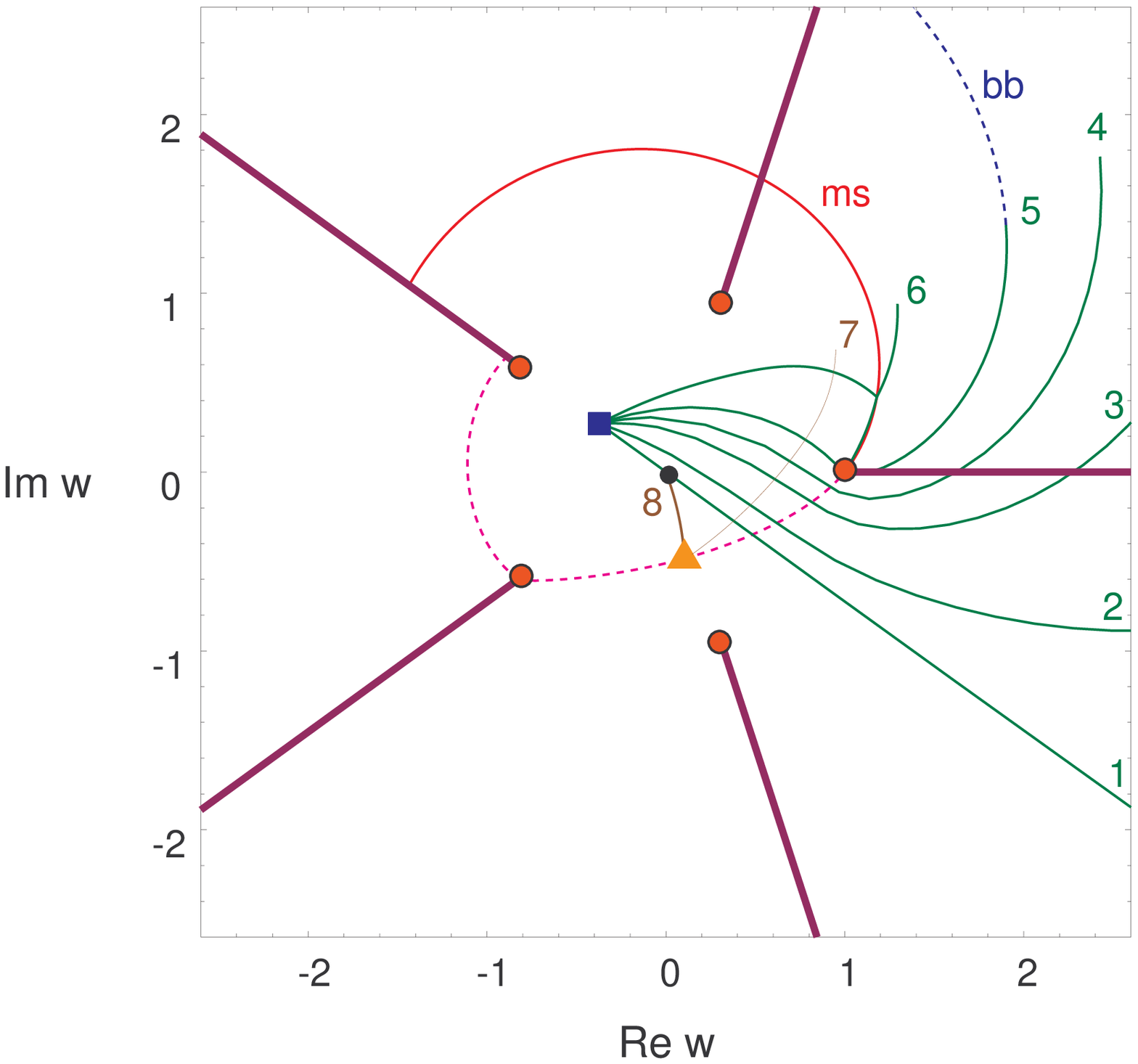,height=8cm}}
\caption{Decay of a BPS state with charge $q=(2,-1,-2,2)$ when
moving from large complex structure $w=\infty$ to the Gepner point
$w=0$ (see section \ref{sec:notconv} for the relation between $w$
and $\psi$). The state decays into $2 (1,0,0,0) + (0,-1,-2,2) = 2
(1,0,0,0) + (-1,0,-4,0)^2$. The green and brown lines $1$ to $8$
are flows with charge $q$, respectively starting at $w_1=\infty$,
$w_2 = 8 \, e^{i \pi/5}$, $w_3=4 \, e^{i \pi/5}$, $w_4=3 \, e^{i
\pi/5}$, $w_5 = 2.35 \, e^{i \pi/5}$, $w_6 = \, 1.6 e^{i \pi/5}$,
$w_7= \, 1.2 e^{i \pi/5}$, and $w_8=0$ (note that the points
$w_1$, $w_2$ and $w_3$ are not in the picture). The blue square is
a regular attractor point, and the orange triangle a crash-zero.
Beyond the basin boundary ``bb'' separating the good and the bad
attractor point, the flow is split. Beyond the marginal stability
line ``ms'', the state has decayed.}\label{MrDecay}}

We now turn to the analysis of some split flow examples. As
explained in section \ref{sec:existencecrit}, a BPS solution
corresponding to a split flow decays when the vacuum moduli are chosen
to lie on the line of marginal stability where the split point is located. Such a decay
does not necessarily mean that the charge disappears completely from the
(supergravity) BPS spectrum: it is perfectly possible that it
still exists in a different realization, for example as a single
flow.

Most interesting are the cases, though, where the charge does
indeed disappear from the BPS spectrum completely when going from
one region of moduli space to the other. Such examples highlight
qualitative differences in the physics associated to one region or
the other.

In fig.\ \ref{MrDecay}, we present an example
of a charge that exists as a BPS state at large radius, but is
(very likely) no longer in the spectrum at the Gepner point. Its
charge in the $\Pi$-basis is $\Gamma=q=(2,-1,-2,2)$. At large
radius $\psi \to \infty$ it is realized as a single center black
hole. In fact, it is a cousin of our lightest BPS black hole state
$(1,0,4,0)$, discussed above, since
$(2,-1,-2,2)=-(1,0,4,0).T[\infty]^3.T[0]^{-1}$. When approaching
the Gepner point along the $\arg \psi = \pi/5$ axis, it gets
transformed into a split flow with one leg on the conifold point
$\psi=1$, by the mechanism of section \ref{sec:multibasin}: two D6
particles are created at $r=r_{ms}$. When continuing further
towards the Gepner point, the line of marginal stability for this
split flow is crossed, and the state decays, by expelling the two
units of $D6$ charge to spatial infinity.

Our numerical analysis strongly suggests that the charge has
disappeared completely from the BPS spectrum at the Gepner point.
It is easy to verify that it does not exist there as a single
flow. Moreover, it clearly does not exist as a split flow at the
``crash point'' where $Z=0$. This is quite obvious from energy
considerations, or alternatively, it can be argued as follows:
there is no room left for the would-be branch running to a split
point (since $|Z|$ must decrease along a flow), and there can be
no branches running away from it, since if a line of
$(\Gamma_1,\Gamma_2)$-marginal stability with
$\Gamma_1+\Gamma_2=\Gamma$ ran through the zero, we would have
simultaneously $Z(\Gamma_1)=-Z(\Gamma_2)$ and $\arg
Z(\Gamma_1)=\arg Z(\Gamma_2)$, hence $Z(\Gamma_1)=Z(\Gamma_2)=0$,
and again no flow room is left, which proves the claim. Now,
moving away from $Z=0$ upstream the flow, towards the Gepner
point, could open up the possibility to have a split flow if $|Z|$
becomes sufficiently large. At the Gepner point, we have $|Z|
\approx 1.20751$. Since our numerical data shows beyond reasonable
doubt that all regular black holes have mass above $M_{min}
\approx 1.250947$ (see fig.\ \ref{spectrumlines} (4)), any split
flow with charge $\Gamma$ starting from the Gepner point could
only have constituent charges with zero attractor mass, i.e.\ pure
$D6$ or $D2-D0$ relatives. However, using the existence criteria
for split flows of section \ref{sec:existencecrit}, we excluded
(with the help of a computer) the existence of $\Gamma \to
\Gamma_1 + \Gamma_2$ split flows with $\Gamma_1$ and $\Gamma_2$
$\IZ_5$-relatives of $k D_6$ and $n D2 + m D0$ charges, with
$|k|,|n|,|m|$ smaller than $25$ (greater charge numbers give
masses that are way too high along the flow under consideration).
A glance at the lower end of the mass spectra in fig.\
\ref{spectrumlines}, keeping in mind the general arguments of
appendix \ref{sec:strings}, shows that this is a quite manageable
task. In principle, it is then still possible that more
complicated split flows with the given charge exist, with more
than two legs of $D6$ or $D2-D0$ type, or with constituents
related to $D6$ and $D2-D0$ through more complicated monodromies
(with consequently significantly longer, hence more massive,
attractor flows). We did not systematically screen those, but
based on the study of a large number of candidates (all with
negative result), we are convinced that it is extremely unlikely
that they would give rise to a valid split flow of the given
charge. Finally, as a check on the above reasoning, we (partially)
verified the nonexistence of a $(\Gamma_1,\Gamma_2)$ split flow
starting at the Gepner point for $\Gamma$ as above, by screening a
set of $6,765,200$ candidate constituent charges
$\Gamma_1=\tilde{q}$ with $-25<\tilde{q}_{2j}<25$, using the
existence criteria of section \ref{sec:existencecrit}, again with
negative outcome.

In conclusion, we predict the existence of a BPS state of charge
$(2,-1,-2,2)$ at large radius that does not exist at the Gepner
point.

\FIGURE[t]{\centerline{\epsfig{file=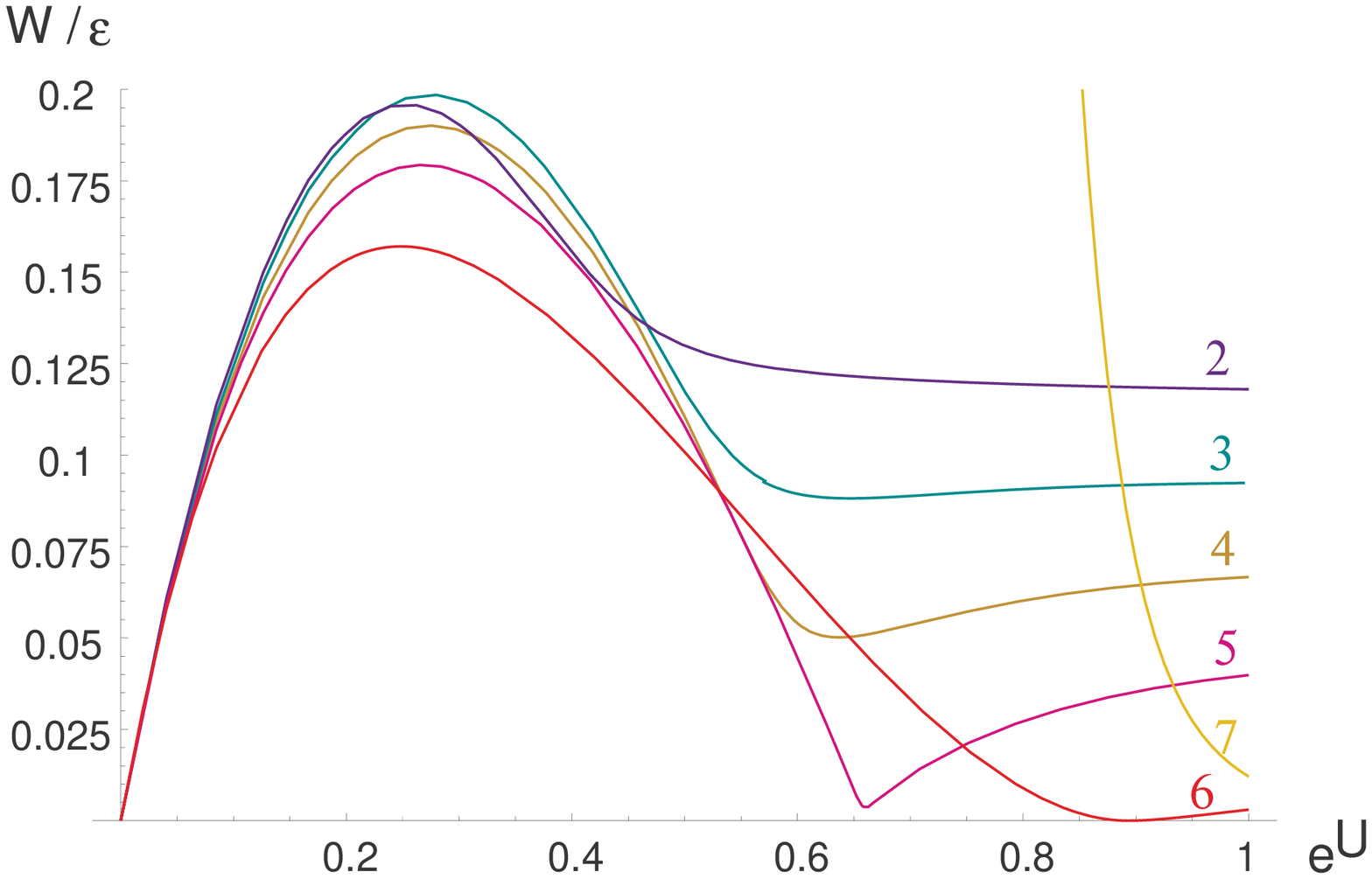,height=7cm}}
\caption{Potentials $W$ for test particle with charge $\epsilon
(1,0,0,0)$ in the backgrounds corresponding to the flows in fig.\
\ref{MrDecay}. As a convenient radial coordinate, we use the
gravitational redshift factor $e^U$.} \label{MrDpot}}

We close this section with a look at the spacetime features of
this solution. An instructive way to understand the stability of
multicenter BPS configurations corresponding to split flows, is
considering the force potential on a test particle of charge
$\epsilon \Gamma_t$ in the given background \cite{branessugra}:
\begin{equation}
 W=2 \, \epsilon \, e^U \, |Z(\Gamma_t)| \,
 \sin^2(\frac{\alpha_t-\alpha}{2})\, ,
\end{equation}
where $\alpha_t = \arg Z(\Gamma_t)$. This potential gives the
excess energy of the configuration over its BPS energy, is
everywhere positive, and becomes zero when $\alpha_t=\alpha$,
explaining why the constituents of such bound states have to be
located at a marginal stability locus in space. Fig.\ \ref{MrDpot}
shows this potential $W$ for a test charge $\epsilon (1,0,0,0)$ in
the background of a spherically symmetric configurations
corresponding to the flows of fig.\ \ref{MrDecay}, nicely
illustrating the appearance of a BPS minimum upon crossing the
basin boundary, and its disappearance upon crossing the line of
marginal stability.

\subsection{An interesting black hole bound state}

\FIGURE[t]{\centerline{\epsfig{file=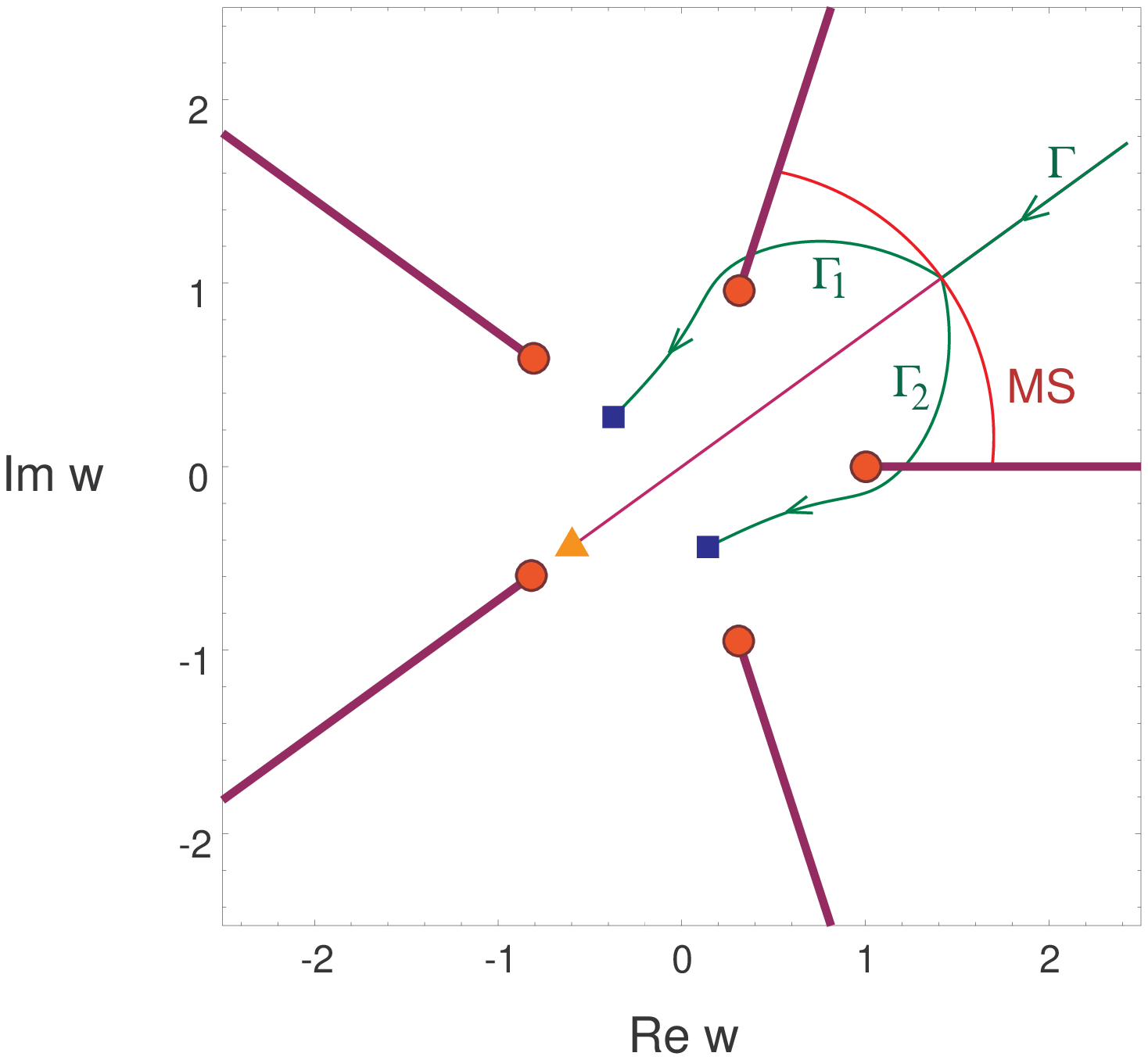,height=8cm}}
\caption{Split flow with regular constituents, which exists at
large complex structure, but not at the Gepner point. The total
charge is $\Gamma=(0,3,9,-8)$, decaying into $\Gamma_2+\Gamma_1 =
(-1,1,4,-1) + (1,2,5,-7) = (1,0,4,0).T[0]^{-1} +
(1,0,4,0).T[0].T[\infty]$. The purple line extending the incoming
branch at the split point is the would-be single flow, crashing on
a regular zero at the orange triangle.}\label{regsplit}}

It is fairly easy to find split flows with two regular black hole
constituents that also have a regular single flow representation.
Finding examples without the latter turns out to be much more
difficult.\footnote{Roughly, one needs an obstruction for smooth
interpolation between the two regular attractor points. In our
example, this obstruction is delivered by the conifold points
``inside'' the flow branches.} One example, with charge
$q=(0,3,9,-8)$, is given in fig.\ \ref{regsplit}.

\FIGURE[t]{\centerline{\epsfig{file=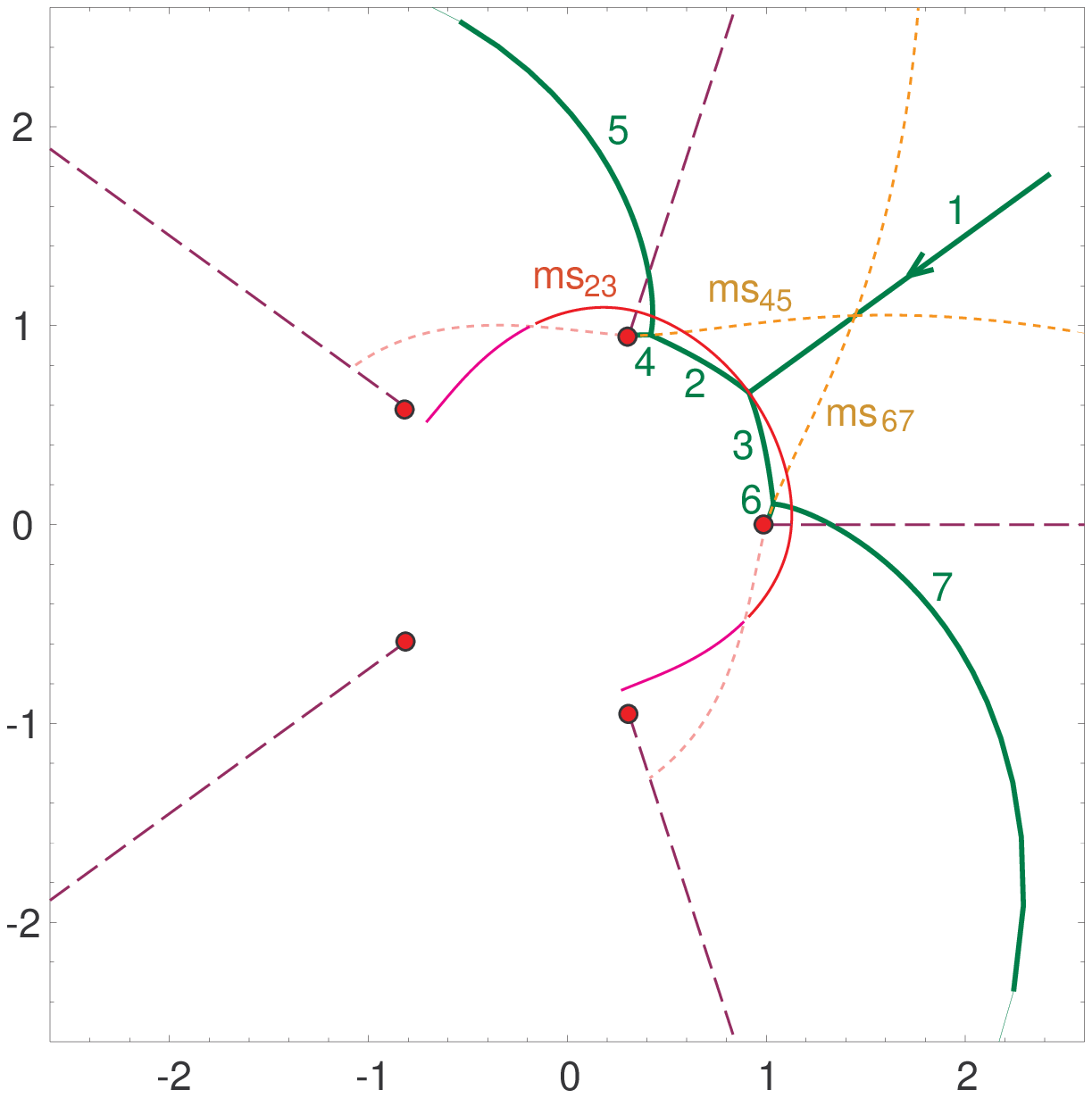,height=9cm}}
\caption{Alternative split flow to fig.\ \ref{regsplit}: the
initial branch (1) of charge $(0,3,9,-8)$ splits at the MS line
$\mathrm{ms}_{23}$ in a branch (2) of charge $(3,3,10,-11) =
(0,0,1,3)^1$ and a branch (3) of charge $(-3,0,-1,3) =
(0,0,-1,2)^{-1}$. Branch (2) then splits on $\mathrm{ms}_{45}$ as
(4): $3(1,0,0,0)^1$ $+$ (5): $(0,0,1,4)$ and similarly (3) on
$\mathrm{ms}_{67}$ as (6): $3 (-1,0,0,0)$ $+$ (7): $(0,0,-1,3)$.
Note that the split flows (245) and (367) can be created from
single flows (of $D0-D2$ type) by the conifold branch creation
mechanism of section \ref{sec:multibasin}.}\label{Mr0398bis}}

However, this example has an alternative split flow realization,
with decay line closer to the Gepner point than the one displayed
in fig.\ \ref{regsplit}. This is shown in fig.\ \ref{Mr0398bis}.
The flow has \emph{four} legs, with none of them corresponding to
a regular black hole constituent: two are of empty hole type
(D6-like), and two of mildly naked type (D0-D2-like).

Again, we did a screening of possible constituents for this
charge, at $w = 2 \, e^{i \pi/5}$, checking $6,765,200$ candidate
charges $\tilde{q}$, with the $\tilde{q}_{2j}$ ranging from $-25$
to $25$. The only two possibilities that came out\footnote{The
screening procedure we used also spits out the second possibility,
even though it has more than two legs. This is because the two
branches in which the incoming branch splits have charge related
to $D2-D0$ by $\IZ_5$ monodromy, and those are automatically put
on the shortlist, without having to exist as single flows (they do
not, in this case).} are those given in fig.\ \ref{regsplit} and
\ref{Mr0398bis}. Therefore, in particular, we expect this charge
to be absent from the BPS spectrum at the Gepner point, with a
somewhat lower level of confidence though (it becomes slightly
less unlikely that more complicated split flows exist).

Incidentally, we did not find any example of a charge that can
\emph{only} be realized as a split flow with exclusively regular
black holes as constituents (i.e.\ like fig.\ \ref{regsplit} but
then without the alternative of fig.\ \ref{Mr0398bis}), but our
search for those was not sufficiently systematic to be conclusive.

\subsection{Multiple basins of attraction}

\FIGURE[t]{\centerline{\epsfig{file=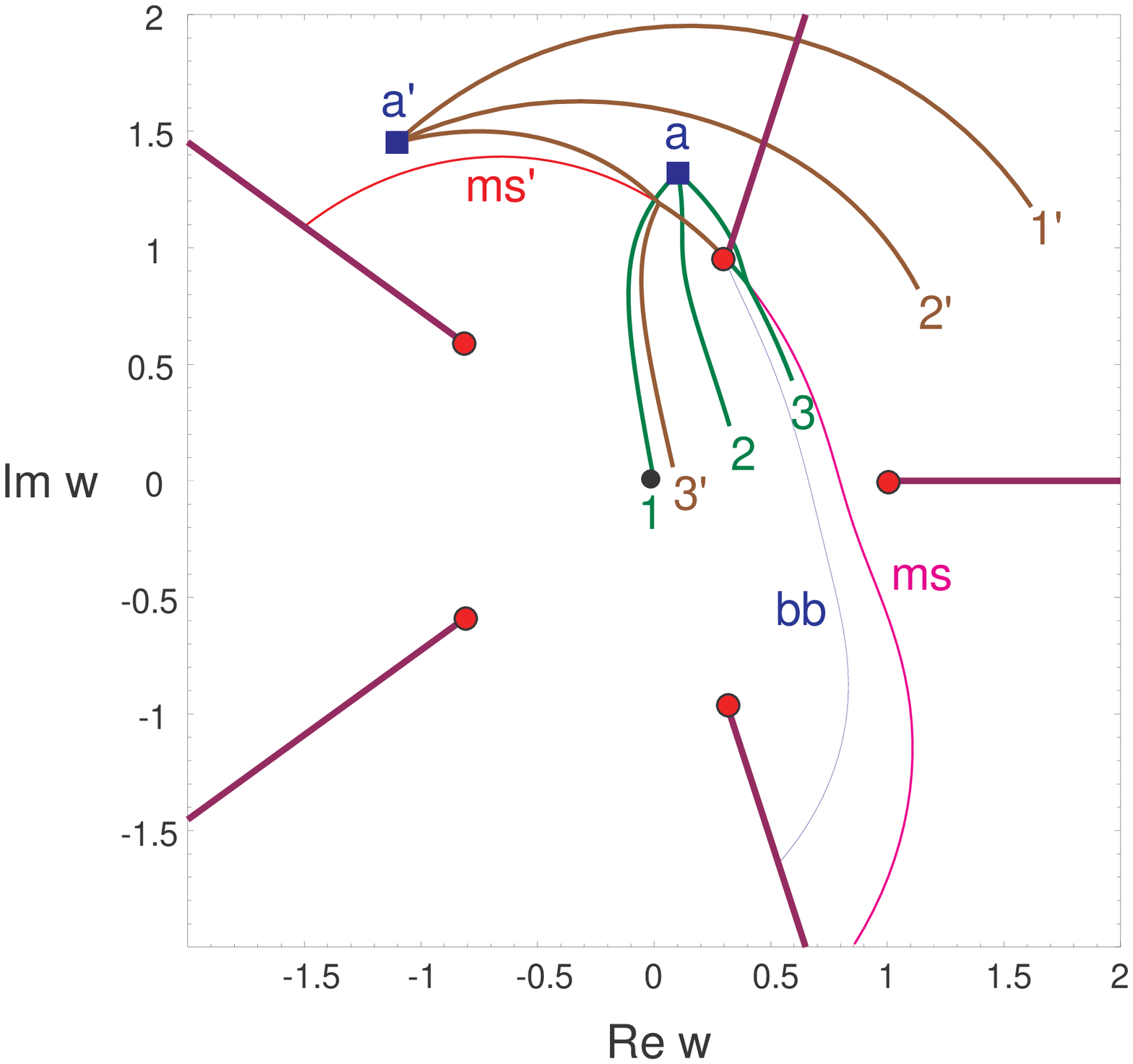,height=7cm}}
\caption{The charge $q=(0,-3,10,18)=(1,-3,-5,-1)^1$ has two
different basins of attraction in the fundamental domain $0 < \arg
\psi < 2\pi/5$, separated by the blue basin boundary labeled
``bb''. Attractor point $a$ has $|Z_*|=6.045$, while attractor
point $a^\prime$ has $|Z_*|=5.603$, so the corresponding near
horizon geometries and entropies of the black holes are not equal.
The green (brown) lines $1,2,3$ ($1',2',3'$) are a continuous
family of flows attracted to $a$ ($a'$), with $3$ and $3'$ the
split flows obtained upon crossing the basin boundary. Those split
flows decay at the marginal stability lines $ms$ resp.\ $ms'$.}
\label{basinsquintic}}

In the previous sections, we have already seen some examples of
multiple basins of attraction induced by the presence of conifold
points. Those were all cases where one of the basins did not have
a good attractor point but a zero instead. The example presented
in fig. \ref{basinsquintic} shows that it is also possible to have
different basins with each a regular attractor point, yielding
different black holes, with different entropies. Recall however
that there is no continuous way to deform the one black hole into
the other (while keeping the BPS property), so there is no
physical consistency problem.

On the full Teichm\"uller space, there will in general be
infinitely many different basins of attraction, corresponding to
the infinitely many ways one can run around the conifold point
copies. This multitude of basins is in strong contrast with the
five dimensional case, where different basins occur much less
generically \cite{multbas}.

\subsection{The D6-D2 system and comparison with the LCSL approximation}

In the large complex structure limit (LCSL), it is possible to
solve explicitly equation (\ref{attreq}) determining the attractor
point \cite{shmakova,M}. The starting point are the asymptotic
expressions (\ref{asymptoticperiods1})-(\ref{asymptoticperiods4})
for the periods, dropping the constant term for $\Pi_{D6}$.
Consider a charge $q=(q_6,q_4,q_2,q_0)$. It is useful to define
the following shifted (nonintegral) charges:
\begin{eqnarray}
 \hat{q}_2 &=& q_2 - \sfrac{11}{2} \, q_4 - \sfrac{25}{12} \, q_6 \\
 \hat{q}_0 &=& q_0 + \sfrac{25}{12} \, q_4 \,.
\end{eqnarray}
>From the results of \cite{shmakova,M}, it follows that the mass
$M_*$ at the attractor point is given by
\begin{equation} \label{M4}
M_*^4=\sfrac{1}{3} \, {\hat{q_2}}^2 \, {q_4}^2 + \sfrac{8}{45} \,
{\hat{q_2}}^3 \, q_6 + 2 \, \hat{q_0} \, \hat{q_2} \, q_4 \, q_6 +
\sfrac{10}{3} \, \hat{q_0} \, {q_4}^3 - {\hat{q_0}}^2 \, {q_6}^2
\, ,
\end{equation}
provided this quantity is positive. The attractor point $t_*$
itself is given by:
\begin{equation}
 t_* = \frac{ (q_4 \, \hat{q_2} - 3 \, q_6 \, \hat{q_0})
 \, + \, 3 \, {M_*}^2 \, i }{5 \, {q_4}^2 + 2 \, q_6 \,
 \hat{q}_2} \, .
\end{equation}
If (\ref{M4}) is negative, the flow crashes at a regular zero and
(\ref{attreq}) does not have a solution $t_*$. Also, the large
complex structure approximation can only be trusted if $t_* \gg 1$
(though we observed pretty good agreement with the full numerical
results for $t_*>1$).

Note that there are no multiple basins of attraction in this
approximation: $t_*$ is unambiguously fixed by the charge. This is
not surprising, since the LCSL approximation is blind for the
presence of the conifold point. Furthermore, as it should, the
expression for $M_*$ is invariant under the monodromy $q \to q
\cdot T[0]$, while the expression for $t_*$ transforms as $t_* \to
t_*+1$.

\FIGURE[t]{\centerline{\epsfig{file=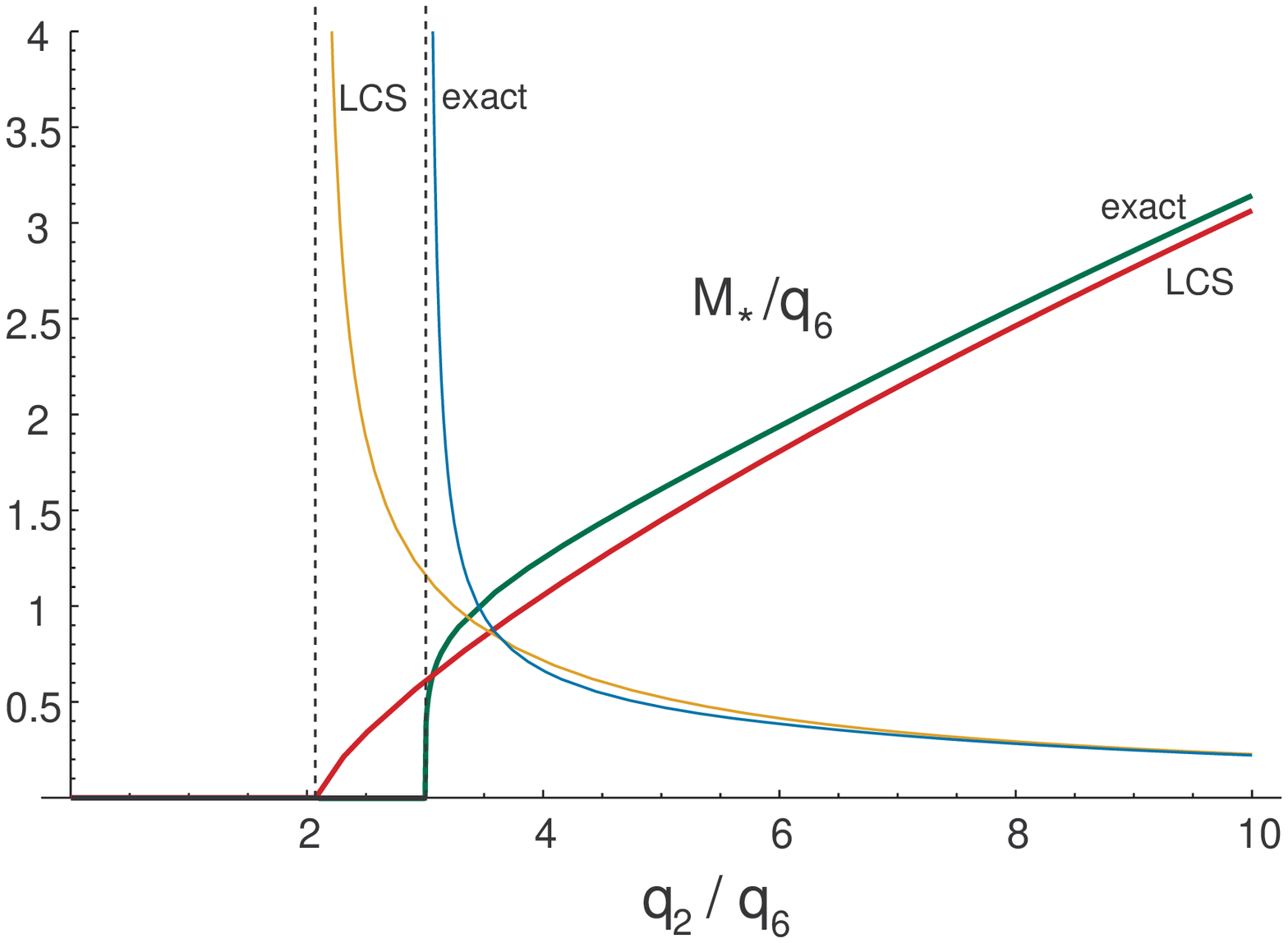,height=7cm}}
\caption{Comparison of numerical results with the LCSL
approximation for charge $(q_6,0,q_2,0)$, $q_6>0$, where $q_2/q_6$
is varied continuously over the $x$-axis. The lines hitting zero
give the relative attractor mass $M_*/q_6$ for the given charge,
with the zero corresponding to the critical point for existence of
a single flow. The lines diverging at those zeros indicate the
mass of the ``dual'' charge defined in (\ref{dualcharge}), with
the divergence corresponding to reaching a boundary of
Teichm\"uller space. Note that the exact $M_*$ curve is much
steeper near its critical point than the LCS curve near its
critical point. This implies that the near-critical single flow
mass spectrum will be much more dilute in the exact than in the
LCS case.}\label{D6D2masses}}

An interesting special case is a pure $D6-D2$ system, i.e.\
$q_4=0$, $q_0=0$. Then the above equations simplify to
\begin{eqnarray}
 \hat{q}_2 &=& q_2 - {\textstyle \frac{25}{12}} \, q_6 \\
 \hat{q}_0 &=& 0 \\
 M_*^4 &=& {\textstyle \frac{8}{45}}\, {q_6}^4 \, (q_2/q_6 - {\textstyle \frac{25}{12}})^3 \label{M4D6D2} \\
 t_* &=& i \, \sqrt{{\textstyle \frac{2}{5}} (q_2/q_6 - {\textstyle \frac{25}{12}})} \, .
\end{eqnarray}
Therefore, according to the LCSL approximation, we need $q_2/q_6
\geq  \frac{25}{12} \approx 2.08333$ for the flow to exist. At the
critical value, we have $t_*=0$. This is a boundary point of the
LCSL Teichm\"uller space, which is a general feature of critical
charges, as can be seen directly from (\ref{attreq}) or
(\ref{attrbound}).

Our numerical results on the other hand indicate that the exact
condition for the existence of a flow is $q_2/q_6 \geq 3$. At this
critical value, we have $q \sim (0,0,2,1)^2$, and the attractor
point is indeed again a boundary point, $\psi=-\infty$. A more
detailed comparison of the exact and LCSL cases is shown in fig.\
\ref{D6D2masses}.

\FIGURE[t]{\centerline{\epsfig{file=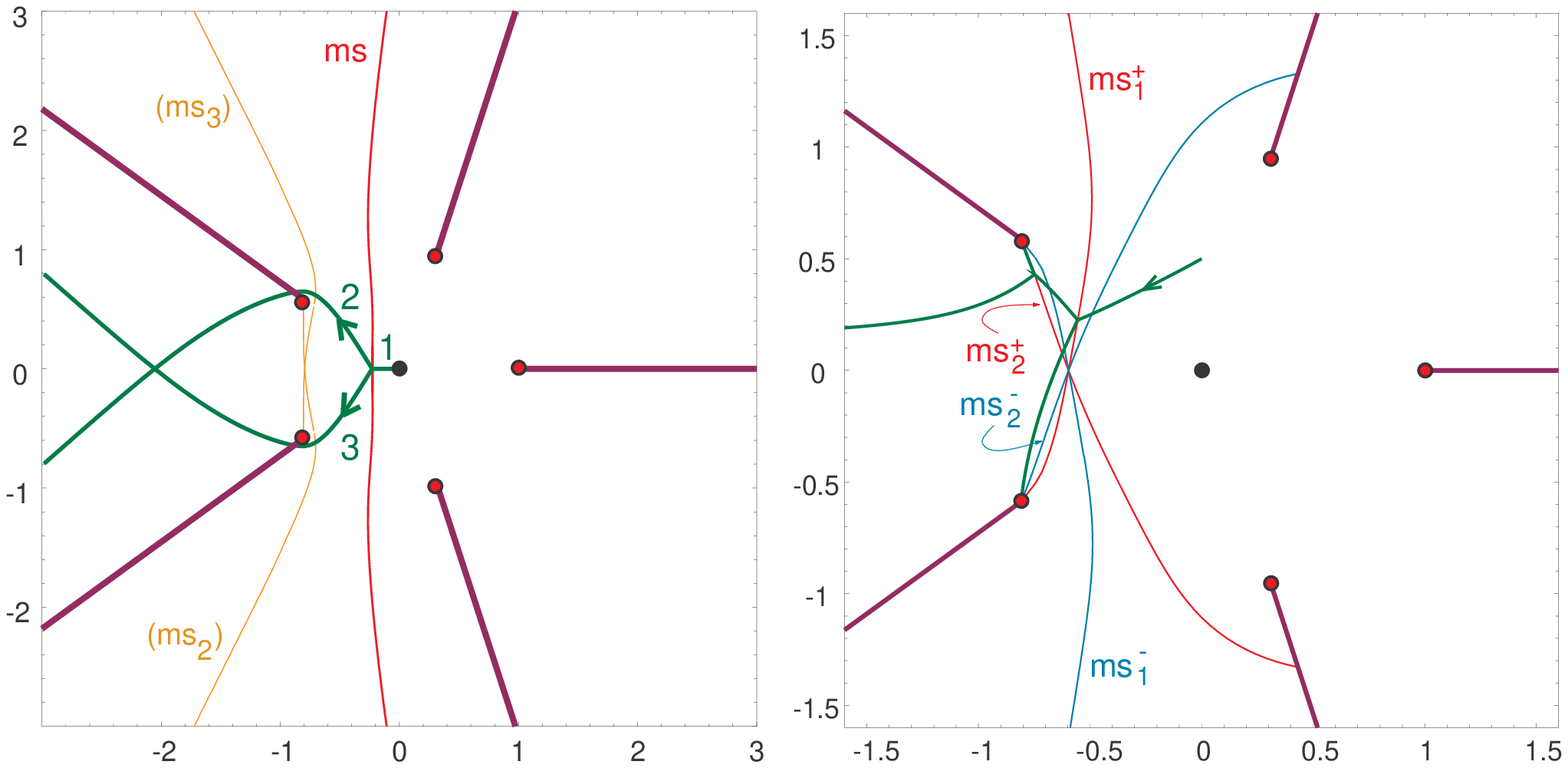,height=8cm}}
\caption{Two possible split flows with charge $q=(3,0,8,0)$, near
the Gepner point. \emph{Left:} two-legged realization, with one
leg (2) of charge $(-2,-2,-7,7) = -(0,0,1,2)^1$ and the other one
(3) of charge $(5,2,15,-7)=(0,0,-1,1)^3$. Note that when the
starting point is moved up sufficiently, branch (3) gets an
additional conifold leg, with marginal stability line $ms_3$.
\emph{Right:} For starting points above the real $w$-axis, we get
an alternative split flow of the form $(3,0,8,0) \to
(1,0,0,0)^3+[(0,0,-2,-3)^{-1} \to (1,0,0,0)^2 + (0,0,-2,-1)^2]$,
with MS lines $ms_1^+$ resp.\ $ms_2^+$. For starting points below
the real axis, one gets a similar but reflected configuration. For
points on the real axis, there is only one split point, with three
outgoing branches.}\label{rdecomp}}

It is plausible that the critical point for the single flow BPS
spectrum is also the critical point for the general flow BPS
spectrum in the LCSL approximation (we did not study this question
systematically though). However, in the exact case, this is
certainly not true. In fact, there is a rich set of split flows
with charge quotient \emph{below} the single flow critical value
$q_2/q_6=3$. We do not know how far below this value one can go
with split flows, but from the discussion at the end of section
\ref{sec:singleflowspectra}, it follows that this is certainly
only a finite amount.

One example is the ``mysterious'' BPS state $|10000\rangle_B$
discovered in \cite{BDLR} and given a split flow interpretation in
\cite{branessugra}. The charge of this state is $(2,0,5,0)$, so
indeed $q_2/q_6 = 2.5 <3$. It exists at the Gepner point, but
decays when moving along the negative $\psi$-axis. The decay
products are $(1,0,0,0)^2$ and $(1,0,0,0)^3$. These are the only
possible decay products that came out of the screening of our
usual $6,765,200$ candidate constituents with
$-25<\tilde{q}_{2j}<25$. Another example is given in fig.\
\ref{rdecomp}. Here we have $q_2/q_6 = 8/3 \approx 2.666667$. At
the Gepner point, there are two possible realizations, again the
only ones resulting from the screening procedure.

Many other examples can be constructed. It would be interesting to
study the spectrum of such ``sub-critical'' split flows
systematically.

\subsection{The monodromy stability problem}
\label{monprob}

\FIGURE[t]{\centerline{\epsfig{file=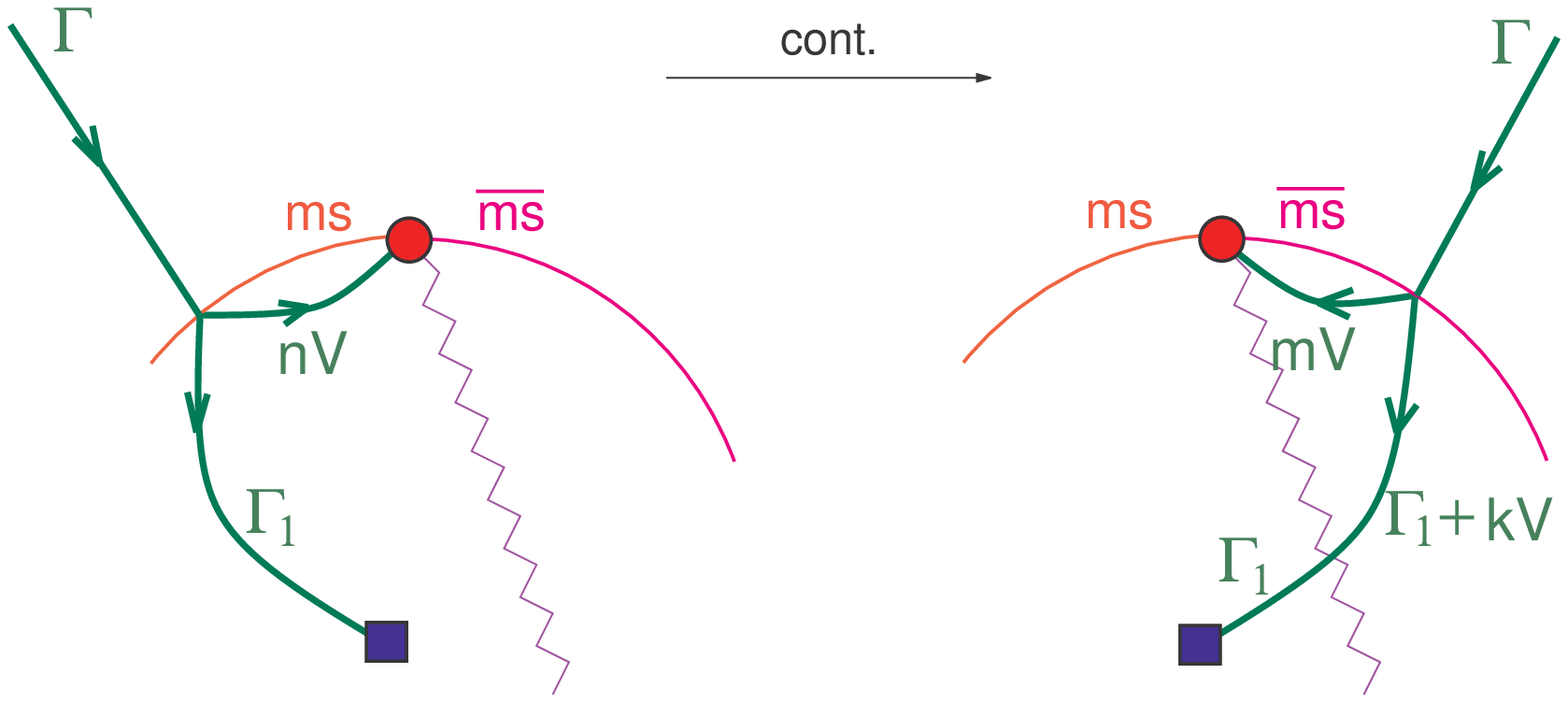,height=6cm}}
\caption{The split flows on the left and the right are related by
continuous variation of the starting point. By charge
conservation, we have $\Gamma=\Gamma_1 + n V = \Gamma_1 + (m+k)
V$, and by the Picard-Lefshetz theorem $k=\langle \Gamma_1,V
\rangle$. The purple wiggly line is the cut corresponding to the
conifold monodromy.  The red line $ms$ is the $( \Gamma_1,n
V)$-marginal stability line; the pink line $\overline{ms}$ is the
``conjugate'' $(\Gamma_1,- n V)$-marginal stability line (as
$Z(V)$ flips sign upon crossing the conifold singularity).
}\label{monodromprobsketch}}

Common ${\cal N}=2$ lore states that BPS states can only decay
when a line of marginal stability is crossed, or, in the split
flow picture, when the incoming branch shrinks to zero size.
However, a paradox arises in some cases, due to monodromy effects.
Consider, as in fig.\ \ref{monodromprobsketch}, a split flow with
one leg on a conifold point, carrying a charge $n V$, where $V$ is
the charge with vanishing mass at the conifold point. The other
leg carries an arbitrary charge $\Gamma_1$. When the starting
point of the flow is continuously moved to the right and all goes
well, the configuration should transform to one with a leg of
charge $m V$, as indicated in the picture, by the mechanism of
section \ref{sec:multibasin}. The integers $n$ and $m$ are related
by
\begin{equation}
 n = m + \langle \Gamma_1,V \rangle \, .
\end{equation}
Consistency requires that the $(\Gamma_1,- n V)$-MS line (labeled
$\overline{ms}$ in the figure) is also a $(\Gamma_1+k V,m V)$-MS
line. This is the case if and only if
\begin{equation} \label{consistency}
 m n \leq 0 \, .
\end{equation}

\FIGURE[t]{\centerline{\epsfig{file=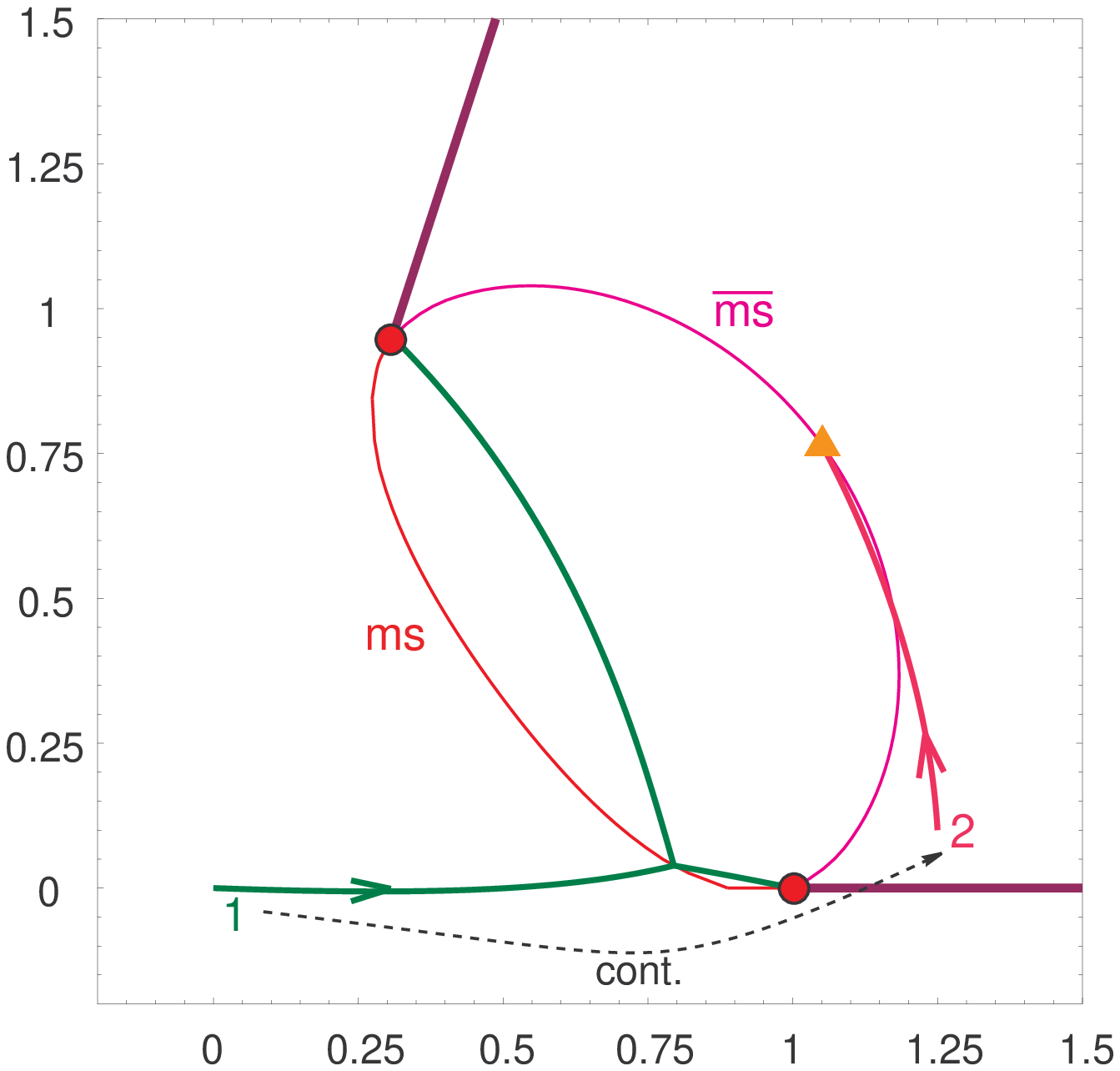,height=7cm}}
\caption{When the starting point of the split flow $(7,1,3,-5) \to
6 (1,0,0,0) + (1,0,0,0)^1$ is continuously transported as
indicated, it ceases to exist: it crashes on the would-be line of
marginal stability. The transport is counterclockwise with respect
to the conifold point (whereas in fig.\ \ref{monodromprobsketch},
it was clockwise), so we have here $m=6$, $n = 6 + \langle
(1,0,0,0)^1,(1,0,0,0) \rangle = 6-5 = 1$, and $n m = 6$, which
does not satisfy (\ref{consistency}).}\label{monodromprob}}

To show that this situation can indeed occur in practice for the
quintic, and to illustrate what happens if the above condition is
not satisfied, we give a specific example not satisfying the
consistency condition in fig.\ \ref{monodromprob}. The figure
shows that when the starting point crosses the basin boundary
induced by the conifold point (i.e.\ the critical incoming flow
passing through that point), we get a flow crashing on the
$\overline{ms}$-line. So the BPS solution ceases to exist. In
general, there might of course still exist other BPS realizations
of the given charge, but we cannot reach these continuously in
this way.

So we are facing a paradox again. One could contemplate the
possibility that the BPS state just decays into other BPS
particles at the basin boundary, but this is not a satisfying
solution: apart from the fact that there is no smooth spacetime
description of such a hypothetical decay, it follows directly from
equation (\ref{integrated}) that a basin boundary, being an
attractor flow, can never be a marginal stability line of mutually
nonlocal charges, and is very unlikely to be a marginal stability
line for suitable mutually local charges.

A more attractive way out, at least at the supergravity level, is
that the BPS state transforms into a non-BPS state. This is
suggested by studying the test particle potential for charge
$\epsilon V$: indeed, upon crossing the basin boundary, the
minimum of the potential gets lifted to a nonzero value, like in
fig.\ \ref{MrDpot}. However, at the quantum level, such a
transition is usually considered unlikely, because a non-BPS
supermultiplet has more states than a BPS multiplet, so to match
the number of states, BPS states should pair up, which would
require a non-generic degeneracy of distinct BPS multiplets. It is
not impossible that this is precisely what happens here, but we do
not know how exactly it would work.

Another possibility is that the configuration was not BPS to begin
with, either because of quantum subtleties, or because of
subtleties arising already at the classical level in the
construction of truly solitonic multicenter supergravity solutions
involving empty hole charge, discussed briefly (but
inconclusively) at the end of section \ref{sec:spliflo}. Support
for this possibility is given by the strong analogy with the
multi-pronged string picture of BPS states in quantum field theory
\cite{threeprong}, where similar spurious multi-pronged strings
arise, which are discarded on the basis monodromy arguments, or by
lifting the strings to M-theory 2-branes. This is the so-called
``s-rule''. In \cite{argyres}, where BPS solutions of ${\cal N}=2$
effective Yang-Mills theories were studied, a very similar problem
as the one discussed in this section was encountered: the s-rule
did not seem to emerge --- or at least not in an obvious way ---
from the analysis of possible solutions.

\FIGURE[t]{\centerline{\epsfig{file=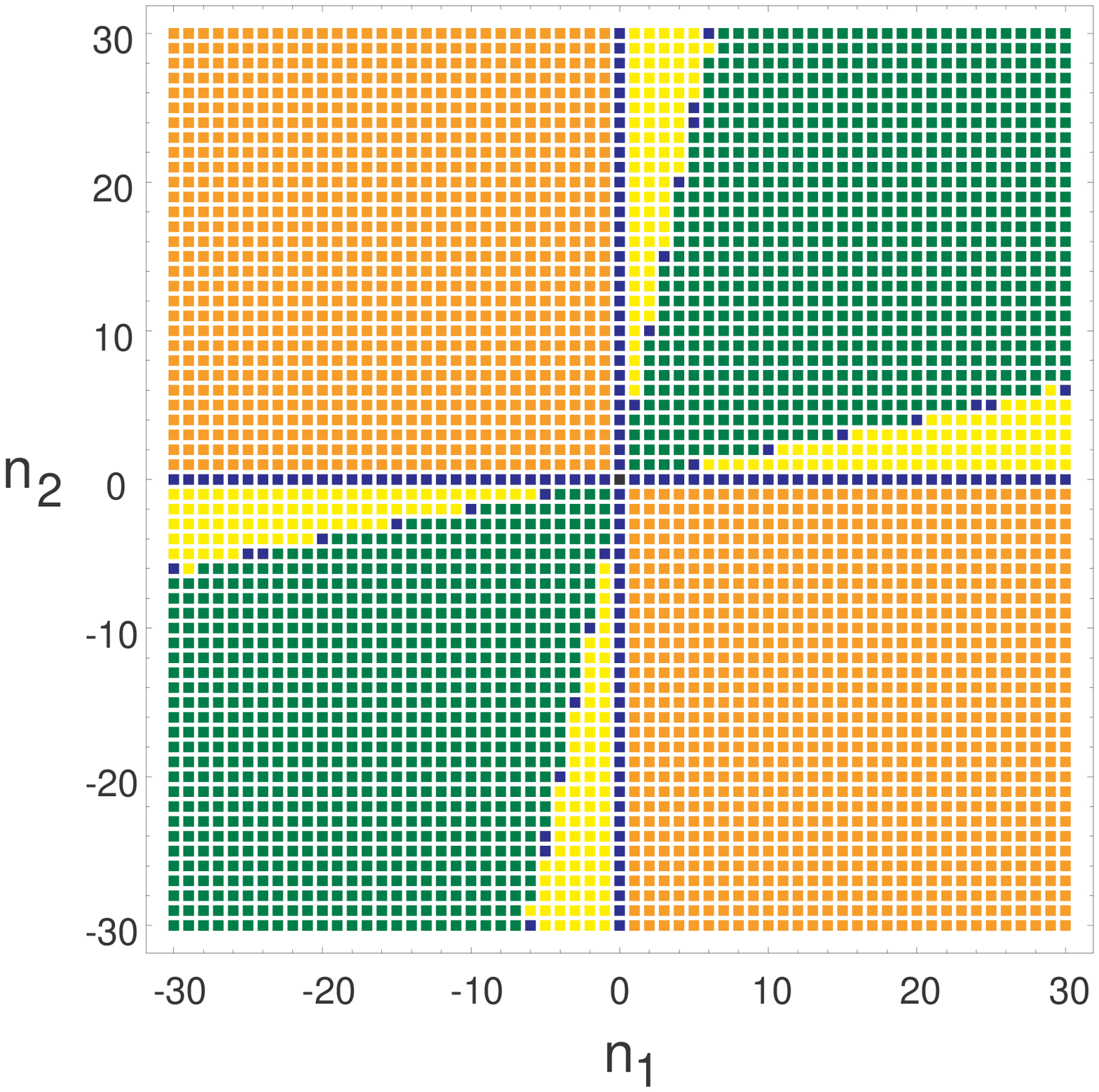,height=8cm}}
\caption{Split flow spectrum at the Gepner side of the MS
ellipsoid for flows with a charge $n_1$ leg on $(1,0,0,0)$ and a
charge $n_2$ leg on $(1,0,0,0)^1$. The orange dots ($n_1 n_2 < 0$)
indicate charges that do not have a split flow realization, the
yellow dots are the charges that have a split flow realization,
but one that is unstable under repeated monodromy around the MS
ellipsoid, and the green and blue dots are the monodromy stable
split flows, with the blue dots the charges related to pure
$(1,0,0,0)$ or the pure $(1,0,0,0)^1$ charge.}\label{conifspec}}

Whatever may be the correct answer, an interesting question
associated for example to the case shown in fig.\
\ref{monodromprob}, is which split flows stay within the spectrum
no matter how many times one encircles the marginal stability
ellipsoid generated by the two conifold point copies considered
there. This gives an infinite set of consistency conditions,
arising from both conifold points. If we write the charge $q$ as
\begin{equation}
 q = n_1 \, (1,0,0,0) + n_2 \, (1,0,0,0)^1 \, ,
\end{equation}
and denote the intersection product of the ``right'' charge with
the ``left'' charge by $\kappa$ (so here, $\kappa=5$ at the Gepner
side of the MS-ellipsoid, and $\kappa=-5$ at the LCS side), then
we found, after a somewhat lengthy analysis that we will not give
here, the following criterion to be in the ``monodromy stable''
part of the spectrum: \emph{either} the charge has to be related by a
(multiple) monodromy around the MS ellipsoid to pure $(1,0,0,0)$
or pure $(1,0,0,0)^1$ charge, \emph{or} it has to satisfy
\begin{equation}
 \frac{1}{\lambda} n_2 \leq n_1 \leq \lambda \, n_2 \, ,
\end{equation}
with
\begin{equation}
 \lambda = \frac{1}{2} (\kappa + \sqrt{\kappa^2-4}) \, .
\end{equation}
As a check, note that in the well-known $SU(2)$ Seiberg-Witten
case, where $\kappa=\pm 2$, this condition indeed reproduces the
BPS spectrum.\footnote{As usual in the supergravity picture, we
include multi-particle states in the spectrum.} In the quintic
case at hand, we find at the Gepner side
$\lambda_G=\frac{1}{2}(5+\sqrt{21}) \approx 4.79129$, and at the
LCS side $\lambda_L = -1/\lambda_G$. The resulting monodromy
stable spectrum at the Gepner side is shown in fig.\
\ref{conifspec}.

It is not inconceivable that the subtleties we see appearing here
are related to the $\IR$-valued gradings of the subobjects of
\cite{DFR,Dcat}. It would be interesting to analyze the states of
this section in that framework.

\subsection{The surface area of empty holes and composites is
bounded from below.}

An intriguing feature of all BPS solutions described here is that
they generically have a minimal size, in the sense that the
surface enclosing the region in which the sources can be localized
has, for a given total charge $\Gamma=q$ but arbitrary moduli
values, always a nonzero minimal area. In view of the holographic
principle, this is perhaps not surprising, but at the level of the
equations and their solutions, this is not obvious at all, since
for instance the \emph{coordinate} radius can be made arbitrarily
small.

The smallest area we have found is that of the black hole with
charge $(1,0,4,0)$ (and its cousins under the duality group):
$A_{\mathrm{min}}=19.664719$.

For regular black holes, the above minimal area statement is of
course obvious. For composites, say a charge $\Gamma_1$ surrounded
by a homogeneous shell of total charge $\Gamma_2 \equiv
(\tilde{q}_6,\tilde{q}_4,\tilde{q}_2,\tilde{q}_0)$, we will prove
this now. Note that in general, one can expect the object to
become minimal in size in the large complex structure limit,
because the flow will become infinitely long then, and longer
flows correspond usually to smaller cores. Therefore, we should in
particular consider this limit and show that the asymptotic area
is finite.

We can assume $(q_6,q_4) \neq 0$ because composites of $D2-D0$
type do not exist in the large radius limit. Define $\ell_1$ and
$\ell_2$ as in (\ref{L1})-(\ref{L2}). Taking the intersection
product of the $\ell_i$ with (\ref{integrated}) gives:
\begin{equation} \label{ZLeq}
 e^{-U} \, \im[e^{-i \alpha} Z(\ell_i)] = \im[e^{-i \alpha}
 Z(\ell_i)]_{\infty} \, .
\end{equation}
Denoting the point where the flow splits by $z_*$, this implies
that the value of $U$ there, $U_*$, is given by:
\begin{equation}
 e^{-U_*} = \frac{\im[ \bar{Z} Z(\ell_1) ]_\infty}{\im [\bar{Z}
 Z(\ell_1)]_*}\, .
\end{equation}
Combining this with equation (\ref{rmsformula}) for $r_*=r_{ms}$
and the form of the metric (\ref{staticmetric}) yields
\begin{equation} \label{Aformula}
 A_* = 4 \pi e^{-2U_*} r_{ms}^2 = \pi \langle \Gamma_1,\Gamma_2
 \rangle^2 \, \frac{1}{\left(\im [e^{-i \alpha} Z(\ell_1)]_*\right)^2}  \, \left(
 \frac{\im(\bar{Z} Z(\ell_1))_\infty}{\im(\bar{Z} Z(\Gamma_2))_\infty}
 \right)^2\, .
\end{equation}
As expected, the only vacuum values of the moduli where our
minimal area statement could go wrong is where some central
charges diverge, that is, in the large complex structure limit
$\psi \to \infty$. However, note that in this limit $\im[e^{-i
\alpha} Z(\ell_2)]_\infty = 0$, because the central charges of the
$D2$ and the $D0$ are zero at large complex structure (due to the
K\"ahler potential factor). Therefore, equation (\ref{ZLeq}) tells
that for all $z$ along the flow, we have $\im[e^{-i \alpha}
Z(\ell_2)]_\infty = 0$; that is, in the limit under consideration,
our flow is just a $(\ell_1,\Gamma)$ (anti-)marginal stability
line, and the point $z_*$ is the intersection of this line with
the $(\Gamma_1,\Gamma_2)$ marginal stability line. So the second
factor in $(\ref{Aformula})$ will converge to a fixed finite
value. As for the third factor, using the asymptotic expressions
for the periods given in appendix \ref{numerics}, a direct
computation gives
\begin{equation}
 \lim_{\psi \to \infty} \, \frac{\im(\bar{Z} Z(\ell_1))_\infty}{\im(\bar{Z} Z(\Gamma_2))_\infty}
 = \frac{q_4^2+q_6^2}{q_6 \tilde{q}_4 - q_4 \tilde{q}_6} \, .
\end{equation}
In conclusion, we find for the area of the core, in the large
complex structure limit:
\begin{equation}
 A_* =  \frac{\pi \langle \Gamma_1,\Gamma_2
 \rangle^2}{\left(\im [e^{-i \alpha}
 Z(\ell_1)]_*\right)^2} \, \left( \frac{q_4^2+q_6^2}{q_6 \tilde{q}_4 - q_4 \tilde{q}_6} \right)^2 \, .
\end{equation}
As an example, one finds for the minimal surface area of the state
$|10000\rangle_B$ (with $q=(2,0,5,0)$) discussed in
\cite{BDLR,branessugra}: $A_* = 43.0607$.

For the empty hole, we can make a similar reasoning. Let us
consider the case of N units of pure D6-brane charge. Taking the
intersection product of (\ref{integrated}) with pure D2-brane
charge resp.\ pure D0-brane charge, we get
\begin{eqnarray}
 e^{-U} \im(e^{-i\alpha} Z_2) &=& \im(e^{-i\alpha} Z_2)_\infty \\
 2 e^{-U} \im(e^{-i \alpha} Z_0) &=& N \tau + 2 \im(e^{-i\alpha}
 Z_0)_\infty \, .
\end{eqnarray}
Taking the moduli at spatial infinity to the large complex
structure limit $\psi=\infty$, this becomes
\begin{eqnarray}
 \im(e^{-i\alpha} Z_2) &=& 0 \\
 2 e^{-U} \im(e^{-i \alpha} Z_0) &=& N \tau \, .
\end{eqnarray}
The first equation implies that the flow coincides with the real
$\psi$-axis, with $\psi>1$, where $Z_6$ and $Z_2$ are positive
imaginary. So the second equation yields for the area of the core
of the empty hole, where the attractor point $\psi=1$ is reached:
\begin{equation}
 A_* = 4 \pi (e^{U_*} \tau_*)^{-2} = \frac{\pi N^2}{(\re Z_0)^2_{\psi
 =1}} \approx 43.91946 \, N^2 \, .
\end{equation}
Again, the finiteness of the area is not trivial, since $\tau_*
\to \infty$ and $e^{U_*} \to 0$ in this limit (as can be seen by
intersecting (\ref{integrated}) with $D4$-charge).

\subsection{Near horizon flow fragmentation}

Consider a regular BPS black hole of charge $\Gamma$. In the
near-horizon limit, or equivalently in asymptotically
$\mathrm{AdS}_2 \times S^2$ space, many of the features discussed
above change quite drastically. The split flow picture still
holds, but now with the starting point at the $\Gamma$-attractor
point. The constant term in \ref{Haha} drops out, and as a result
the constraint on the source positions becomes, instead of
(\ref{distconstr}),
\begin{equation} \label{distconstrAdS}
\sum_{q=1}^N \frac{\langle \Gamma_p,\Gamma_q
\rangle}{|\sx_p-\sx_q|} = 0 \, .
\end{equation}
Consequently, multicenter configurations are also generically
allowed for mutually local charges. All this implies we get in
general a plethora of possible realizations of the system as a
multicenter BPS solution, generalizing the AdS-fragmentation
phenomenon described in \cite{adsfrag}. Note however that from the
point of view of an observer far away from the black hole, all
these different solutions will be indistinguishable. If a counting
of the statistical entropy within this low energy framework were
possible, the number of different ways of ``fragmenting'' the flow
in constituents would presumably give a significant contribution
to the entropy.

\section{Conclusions}\label{sec:conclusions}
\setcounter{equation}{0}

We discussed a variety of (at least suggestive) results on the
stringy BPS spectrum of type II Calabi-Yau compactifications that
can be obtained in the framework of BPS supergravity solutions and
their associated split flows, and illustrated this in detail for
the example of type IIA theory on the quintic. Among the specific
predictions we obtained for this example (with various degrees of
confidence) are the following:
\begin{itemize}
 \item There are infinitely many BPS states at any point in moduli
 space. In particular, the set of rational boundary states at
 the Gepner point constructed in \cite{BDLR} is only a small
 fraction of the total spectrum.
 \item The lightest possible regular BPS black hole with charge $N q$
 (in the $\Pi$-basis of section \ref{sec:quinticperiods}) has
 mass equal to $N M_{min}$, with $M_{min} \approx 1.250947$,
 $q = (1,0,4,0)$, and lives in a vacuum with the value at spatial infinity of $\psi \approx 0.375603$.
 From equation (\ref{M4D6D2}), it follows that about $15.44\%$ of
 its mass there is due to worldsheet instanton corrections.
 \item The mass spectrum of BPS states is generically discrete and without
 accumulation points.
 \item A BPS state with charge $q=(2,-1,-2,2)$ exists at large
 radius, but not at the Gepner point. Its decay products are two
 pure $D6$ branes and a cousin of our ``lightest black hole'' particle with charge
 $q=(0,-1,-2,2)=-(1,0,4,0)^2$ (see section \ref{sec:notconv} for the
 superscript notation).
 \item BPS states have usually several distinct low energy
 realizations (e.g.\ as a single center black hole, and as one or
 more multicenter configurations with mutually nonlocal
 components). Each possible composite has its own marginal
 stability line and decay products. The decay products of a state are therefore
 in general \emph{not} fixed by its charge alone.
 \item The charge $q=(0,3,9,-8)$ has at large radius a realization as a composite
 of two regular black holes, of charge $(-1,1,4,-1)=(1,0,4,0) \cdot T[0]^{-1}$ and
 $(1,2,5,-7)=(1,0,4,0) \cdot T[0] \cdot T[\infty]$, but not as a
 single center black hole. It has an alternative realization with
 four non-black hole constituents related to $D6$ and $D2-D0$ type
 charges as shown
 in fig.\ \ref{Mr0398bis}. At $w=2 e^{i \pi/5}$ (and downstream
 the attractor flow from there), these are probably the only two
 possible realizations. The second one decays closer to the Gepner
 point than the first one.
 \item Multiple basins of attraction, with different corresponding
 black hole entropies, are a generic --- and perfectly consistent ---
 feature in the presence of conifold points.
 \item Below the critical value $q_2/q_6=3$ (but not too much) where $D6-D2$ charges
 can no longer be represented as black holes, there is a rich set
 of composite realizations of these charges. They all involve
 constituents related to charges of $D6$ and $D2-D0$ type.
 \item ``Monodromy-stable'' bound states of the D6 $(q=(1,0,0,0))$ and its cousin
 $q=(1,0,0,0)^1$, at the Gepner point, are given by fig.\
 \ref{conifspec}. An example is (a $\IZ_5$-relative of)
 the state $|10000\rangle$ of \cite{BDLR}, with
 charge $q=(2,1,3,-5)=(2,0,5,0)^3=(1,0,0,0)+(1,0,0,0)^1$.
 \item No matter how one tunes the moduli, one can never localize
 $N$ sources in an area less than about $20 N^2$ in Planck units,
 whether the object is a black hole or not (at least for BPS
 solutions).
 \item On the horizon of a BPS black hole, there is a large enhancement
 of possibilities of multicenter configurations.
\end{itemize}

It is actually quite surprising that the split flow picture, if
taken seriously, has so much predictive power on the BPS spectrum:
a priori, it would seem that an enormous amount of possible split
flows would be allowed, much more than the possible decays allowed
by the microscopic picture, but --- at least for charges with low
mass --- this turns out not to be the case; for instance the fact
that out of the $6,765,200$ candidate constituents we screened for
the above discussed charges, only one or two were actually valid,
is quite remarkable.

Notice however that we have made quite a big leap in faith in
accepting this really as a trustworthy prediction. Indeed, we have
made our arguments for low charge numbers, for which supergravity
cannot necessarily be trusted, and while any low charge solution
can always be promoted to a high charge solution by simply scaling
everything up with a large factor $N$, the opposite is
$\emph{not}$ true. In particular, this means that in screening the
possible constituents of the charges under consideration, we were
certainly not screening all possible constituents of its large $N$
counterpart. So strictly speaking, the arguments for nonexistence
of other split flows for e.g.\ $q=(2,-1,-2,2)$ apart from the ones
we presented, have little physical foundation in low energy
supergravity itself. However, since a large part of the argument
relies purely on energy conservation considerations (e.g.\ the
fact that the state at the Gepner point is too light to decay in
BPS states corresponding to regular flows), it is not entirely
unfounded. And more importantly, it cannot be denied that the
split flow picture actually \emph{works}.\footnote{This is not
something that follows just from this paper; it is already the
case for low charges when one only considers single flows, as in
the examples of \cite{M}.} A natural conjecture would therefore be
that this picture should also arise somehow from microscopic
considerations, like it does in the description of QFT BPS states
as (possible multi-pronged) strings \cite{BPS37,threeprong}.
Clearly it would be very interesting if this were indeed the case.
The idea is not that wild though, since the basic structures
underlying BPS objects quite universally tend to be valid in a
wide range of regimes, though their interpretation can vary
considerably.

Another loose end is the monodromy stability (or ``s-rule'')
problem discussed in section \ref{monprob}. Some input from the
microscopic picture, or perhaps a deeper analysis of full
multicenter solutions involving $D6$-like charges, could resolve
this puzzle.

In conclusion, we believe we have convincingly demonstrated that,
while it is probably not the ideal device to get insight in the
underlying organizing structures, the split flow picture can
nevertheless provide valuable information, and some quite concrete
intuition, on the problem of BPS spectra of type II Calabi-Yau
compactifications at arbitrary moduli values.


\acknowledgments

We would like to thank Neil Constable, Mike Douglas, Tomeu Fiol,
Juan Maldacena, Greg Moore, Rob Myers, Christian R\"omelsberger
and Dave Tong for useful discussions. This work was supported in
part by DOE grant FG02-95ER40893.


\appendix

\setcounter{equation}{0}

\section{(Split) flows as geodesic strings and discreteness of
the spectrum} \label{sec:strings}

An interesting link, useful to give some intuition for the
spectrum, can be made between (split) attractor flows and the
``7/3/1''-brane picture of BPS states in rigid ${\cal N}=2$
quantum field theories \cite{BPS37,threeprong}. This comes from
the observation \cite{branessugra} that attractor flows can be
considered to be geodesic ``strings'' in moduli space. Split flows
can similarly be interpreted as geodesic multi-pronged strings.
This follows from the fact that attractor flows in moduli space
are minima of the action\footnote{In a suitable rigid QFT limit of
the Calabi-Yau compactification, this reduces precisely to the
string action considered in \cite{BPS37,threeprong}. In this case,
the strings can be interpreted as genuine IIB strings stretched
between certain D-branes in spacetime.}
\begin{equation} \label{stringaction}
 S=|Z_*| + \int \sqrt{V} ds \, ,
\end{equation}
where the startpoint of the string is kept fixed at the vacuum
moduli, $Z_*$ is $Z(\Gamma)$ evaluated at the free endpoint(s) of
the string, $V = 4 g^{a\bb}
\partial_a |Z| {\bpartial}_{\bb} |Z|$, and $ds$ is the line element on
moduli space: $ds^2=g_{a\bb} dz^a d\bz^{\bb}$. Requiring $\delta
S=0$ for variations of the free endpoint fixes the latter to be
located at the attractor point of $\Gamma$. The mass
$|Z(\Gamma)|_{\mathrm{vac}}$ of the BPS supergravity solution
equals the minimal value of the action $S$.

This picture makes it plausible that in any finite region $F$ of
moduli space (or, more precisely, its covering Teichm\"uller
space), away from singularities, the mass spectrum of (split)
flows (and therefore of BPS supergravity solutions) is discrete
without accumulation points and at most finitely
degenerate.\footnote{Note that this is a priori not obvious, since
the set of {\em candidate} BPS masses \{ $|Z(\Gamma)|$ \} is dense
in $\IR^+$, and the set of regular attractor points is generically
dense in moduli space. This problem did not arise in the QFT case
studied in \cite{BPS37,threeprong}, because the only attractor
points there are located on singularities, which do not form a
dense set.} To see this, first note that at a regular
$\Gamma$-attractor point, equation (\ref{attreq}) implies
\begin{equation} \label{attrbound}
 |Z(\Gamma)|_* \geq \frac{1}{2} \frac{|\langle \Gamma',\Gamma
 \rangle|}{|Z(\Gamma')|} \, ,
\end{equation}
for any $\Gamma'$. If the attractor point is in or not too far
away from our finite singularity-free region $F$, $|Z(\Gamma')|$
can be bounded from above, and therefore the first term in
(\ref{stringaction}) will be bounded from below. Furthermore,
since the right hand side of (\ref{attrbound}) is proportional to
the charge $\Gamma$, this bound should grow roughly proportional
to the ``magnitude'' of the charge $\Gamma$. On the other hand, if
the attractor point is far away from the region $F$ (such that
$|Z(\Gamma')|$ can no longer be bounded), the attractor flow going
to that point will be long, and consequently the second term in
(\ref{stringaction}) will be large, or at least bounded from
below. Again, this term will scale roughly proportional to
$\Gamma$.

This makes it plausible that the spectrum will indeed be discrete
and without accumulation points, something that is also strongly
supported by the numerical data we obtained for the quintic. Of
course, a rigorous proof would require a much more lengthy
analysis, but we will not try this here.

\section{Precise expressions for the quintic periods}\label{numerics}

In this appendix, to facilitate reproduction and extension of our
numerical explorations by the interested reader, we will give the
detailed expressions for the quintic periods in terms of the
pre-defined Meijer functions of the \emph{Mathematica} software
package. This is not entirely trivial, since the presence of
monodromies make these definitions convention-dependent. We will
use \emph{Mathematica} syntax to denote the Meijer functions.

Define
\begin{equation}
 c=\frac{1}{\Gamma(\frac{1}{5}) \Gamma(\frac{2}{5}) \Gamma(\frac{3}{5}) \Gamma(\frac{4}{5})}
 \, \, \, \, \, ,
\end{equation}
\begin{eqnarray}
 U_0^-(z) &=& c \,
 \MG[\{\{\sfrac{4}{5},\sfrac{3}{5},\sfrac{2}{5},\sfrac{1}{5}\},\{\}\},\{\{0\},\{0,0,0\}\},-z]\\
 U_1^-(z) &=& \sfrac{c}{2 \pi i}
 \MG[\{\{\sfrac{4}{5},\sfrac{3}{5},\sfrac{2}{5},\sfrac{1}{5}\},\{\}\},\{\{0,0\},\{0,0\}\},z]\\
 U_2^-(z) &=& \sfrac{c}{(2 \pi i)^2}
 \MG[\{\{\sfrac{4}{5},\sfrac{3}{5},\sfrac{2}{5},\sfrac{1}{5}\},\{\}\},\{\{0,0,0\},\{0\}\},-z]\\
 U_3^-(z) &=& \sfrac{c}{(2 \pi i)^3}
 \MG[\{\{\sfrac{4}{5},\sfrac{3}{5},\sfrac{2}{5},\sfrac{1}{5}\},\{\}\},\{\{0,0,0,0\},\{\}\},z] \, ,
\end{eqnarray}
and
\begin{eqnarray}
 U_0^+(z) &=& U_0^-(z)\\
 U_1^+(z) &=& U_1^-(z) + U_0(z) \\
 U_2^+(z) &=& U_2^-(z) \\
 U_3^+(z) &=& U_3^-(z) + U_2^-(z) \, .
\end{eqnarray}
Then the period basis $\{U_j(z)\}_j$ of section
\ref{sec:quinticperiods} is given by $U_j(z)=U_j^-(z)$ if $\im z <
0$ and $U_j(z)=U_j^+(z)$ if $\im z > 0$.

\emph{Mathematica} evaluation of the general Meijer function is
rather slow --- too slow in fact to do interesting calculations in
a reasonable time on a 500 MHz Pentium III. The process can be
sped up enormously by first computing a lattice of values of the
periods and approximating the periods by an interpolating
function. Because the period functions are quite well behaved,
this can be done with acceptable loss in accuracy. Of course, only
a finite region of the $\psi$-plane can be covered with a finite
lattice of evaluation points, but for large $\ln \psi$, the
polynomial asymptotic expressions for the periods can be used:
\begin{eqnarray} \label{asymptoticperiods1}
 \Pi_{D6} &\approx& -{\sfrac{5}{6}} \, t^3 - {\sfrac{25}{12}} \, t +
 {\sfrac{200 \, \zeta(3)}{(2 \pi)^3} } \, i \\
 \Pi_{D4} &\approx& - {\sfrac{5}{2}} \, t^2 - {\sfrac{11}{2}} \, t +
 {\sfrac{25}{12}} \\
 \Pi_{D2} &\approx& t \\
 \Pi_{D0} &\approx& 1 \, ,
 \label{asymptoticperiods4}
\end{eqnarray}
where $t \approx \sfrac{5 i}{2 \pi} \ln(5 \psi)$. This gives for
the K\"ahler potential:
\begin{equation}
 e^{-{\cal K}} \approx \sfrac{20}{3} (\im t)^3 \, .
\end{equation}


\newcommand{\dgga}[1]{\href{http://xxx.lanl.gov/abs/dg-ga/#1}{\tt
dg-ga/#1}}
\renewcommand\baselinestretch{1.08}

\end{document}